# Widespread remote introgression in the grass genomes

Yujie Huang[1], Shiyu Zhang[1], Hanyang Lin[2], Chenxu Liu[3], Zhefu Li[1], Kun Yang[1], Yutong Liu[1], Linfeng Jin[1], Chuanlong Lu[3], Yuan Cheng[3], Chaoyi Hu[4], Huifang Zhao[1], Guoping Zhang[1], Qian Qian[5], Longjiang Fan[1,5,*], Dongya Wu[1,6,*]

[1] Institute of Crop Science, Zhejiang University, Hangzhou, China

[2] School of Advanced Study, Taizhou University, Taizhou, China.

[3] State Key Laboratory for Managing Biotic and Chemical Threats to the Quality and Safety of Agro-Products, Vegetable Research Institute, Zhejiang Academy of Agricultural Sciences, Hangzhou, China

[4] Hainan Institute, Zhejiang University, Sanya, China

[5] Yazhouwan National Laboratory, Sanya, China

[6] Center for Evolutionary & Organismal Biology, Zhejiang University, Hangzhou, China

* E-mail: fanlj@zju.edu.cn to L.F.; wudongya@zju.edu.cn to D.W.

## Abstract

Genetic transfers are pervasive across both prokaryotes and eukaryotes, encompassing canonical genomic introgression between species or genera and horizontal gene transfer (HGT) across kingdoms. However, DNA transfer between phylogenetically distant species, here defined as remote introgression (RI), has remained poorly explored in evolutionary genomics. In this study, we present RIFinder, a novel phylogeny-based method for RI event detection, and apply it to a comprehensive dataset of 122 grass genomes. Our analysis identifies 622 RI events originating from 543 distinct homologous genes, revealing distinct characteristics among grass subfamilies. Specifically, the subfamily Pooideae exhibits the highest number of introgressed genes while Bambusoideae contains the lowest. Comparisons among accepted genes, their donor copies and native homologs demonstrate that introgressed genes undergo post-transfer localized adaptation, with significant functional enrichment in stress-response pathways. Notably, we identify a large Triticeae-derived segment in a Chloridoideae species *Cleistogenes songorica,* which is potentially associated with its exceptional drought tolerance. Furthermore, we provide compelling evidence that RI has contributed to the origin and diversification of biosynthetic gene clusters of gramine, a defensive alkaloid chemical, across grass species. Collectively, our study establishes a robust method for RI detection and highlights its critical role in adaptive evolution.

# Introduction

DNA transfer across species is pervasive across taxonomic groups, acting as a key driver of evolutionary diversification. Previous studies have highlighted the crucial roles of genomic introgression or gene flow in facilitating adaptive evolution (Chen et al., 2023; Edelman et al., 2019a; Ma et al., 2022; Peccoud et al., 2017; Zhao et al., 2023; Li et al., 2022a). For example, introgression of the *EPAS1* allele from archaic Denisovan into modern humans confers high-altitude adaptation in Tibetan populations (Huerta-Sánchez et al., 2014). A recent study reported a remarkable case of introgression event involving the Y sex chromosome between deeply divergent primate lineages in guenons (Jensen et al., 2024). Genomic introgression, through hybridization and successive back-crossing, has become a fundamental strategy in modern crop breeding to introduce beneficial alleles from related wild species into elite varieties. Horizontal gene transfer (HGT), another form of genetic exchange, typically occurs through asexual mechanisms (e.g. parasitism and symbiosis) (Keeling & Palmer, 2008; Soucy et al., 2015). Although relatively rare in eukaryotes, well-documented cases of HGT from prokaryotes have played pivotal roles in shaping adaptive traits in recipient lineages (Etten and Bhattacharya, 2020; Keeling, 2024; Keeling and Palmer, 2008). For instance, acquiring genes from *Listeria* influences male courtship behavior toward females in lepidoptera insects (Li et al., 2022b). The whitefly (*Bemisia tabaci*), a major agricultural pest, has incorporated *BtPMaT1* from its host plants to neutralize plant-derived toxins (Xia et al., 2021). Moreover, gene transfer via physical contacts, such as between parasitic plants and their hosts via haustoria, is recognized as a form of HGT (Yoshida et al., 2010).

Through an extensive review of 5,318 studies on DNA transfer published over the past five years (**Supplementary Fig. 1**), we found that HGT events are predominantly associated with inter-kingdom transfers, whereas genomic introgression is almost exclusively limited to gene flow between closely related taxa (e.g. congeneric species). This dichotomy leaves a conceptual gap regarding the possibility of genetic exchange between distant lineages with moderate phylogenetic distance. Therefore, to clearly differentiate them from these established mechanisms, we herein define such instances of DNA exchange as remote introgression (RI).

The grass family (Poaceae) is among the most vital plant families for humans, serving as the principal source of calories for global populations. Grass species are generally divided into two major evolutionary lineages that diverged > 80 million years ago (mya): the PACMAD lineage, which includes Panicoideae (PAN), Arundinoideae, Chloridoideae (CHL), Micrairoideae, Aristidoideae, and Danthonioideae subfamilies. and the BOP lineage, including Bambusoideae (BAM), Oryzoideae (ORY), and Pooideae (POO) subfamilies (Huang et al., 2022). Several cases of gene transfer have occurred across subfamilies, despite the deeply evolutionary divergence. For example, a fragment containing five genes related to abiotic stress response was reported to transfer from *Panicum* (PAN) to wild barley species (*Hordeum*, POO) (Mahelka et al., 2021, 2017). Likewise, seven *Bx* genes in Triticeae involved in the biosynthesis of DIMBOA (2,4-dihydroxy-7-methoxy-1,4-benzoxazin-3-one), were acquired via RI from ancestral Panicoideae (Wu et al., 2022b). Another compelling RI case involves the transfer of momilactone A biosynthetic genes from wheat (*Triticeae*) to rice (*Oryza*), enabling the production of this allelopathic defense compound in a distantly related lineage (Wu et al., 2022). Although these events were previously described as horizontal or lateral transfers, we categorize them herein as RI to clearly distinguish them from canonical HGT.

Gene evolutionary histories are frequently obscured by progresses such as duplication, loss, transfer, and incomplete lineage sorting (ILS), all of which contribute to gene tree discordance and reticulate evolution (Edwards et al., 2007; Steenwyk et al., 2023). Detecting genomic introgression and HGT under these conditions relies primarily on site pattern analysis and phylogenetic inference. Classic methods such as *D*-statistics (ABBA-BABA test) or *f*-statistics (e.g. *f*3 and *f*4) are widely used. While site-pattern-based methods remain effective for detecting recent gene flow among closely related lineages, their sensitivity declines sharply as phylogenetic distance increases. Moreover, their design limitations, typically confined to four or five taxa, render them ill-suited for identifying RI across broader evolutionary scales, especially in large, complex phylogeny. Phylogeny-based methods (e.g. QuIBL) leverage gene tree topologies and branch length distributions to distinguish introgression from ILS (Edelman et al., 2019b), but they are generally restricted to analyses involving three or four taxa. Similarly, PhyloNet integrates multiple gene trees to infer phylogenetic networks and hybridization events (Than et al., 2008). Collectively, these methods are poorly adapted to uncovering RI across deep evolutionary timescales. With the increasing availability of high quality genomes, the discovery of hidden evolutionary relationships among distantly related species necessitates the development of novel algorithms tailored for large-scale evolutionary studies.

In this study, we aim to establish a comprehensive workflow for identifying RI events across large-scale genomic datasets and systematically examine their prevalence and adaptive significance. To achieve this, we introduce a modular tool, Remote Introgression Finder (RIFinder), specifically designed for RI detection. Using RIFinder, we uncover widespread RI events across 122 grass genomes and characterize the structural and functional features of RI-associated genes. Notably, we report a ~30-Kbp RI segment potentially involved in drought tolerance of the Eurasia desert plant *Cleistogenes songorica*, and elucidate the complex evolutionary trajectory of defense-related gramine biosynthetic gene clusters.

# Results

## Inferring remote introgression events between phylogenetically distant lineages using RIFinder

To efficiently and high-throughput identify RI events among hundreds of genomes, we developed RIFinder, a novel analytical workflow (**Fig. 1a**). It consists of three stepwise modules: phylogenetic inference of homologous groups, RI signal detection, and statistical significance testing (**Supplementary Fig. 2**). The aim of RIFinder is to detect the gene-species phylogeny conflicts among evolutionarily distant species (inter-clade). Firstly, species-level taxonomic clades are defined to construct a reference phylogeny, serving as the backbone for comparing gene-level phylogeny. Protein sequences from multiple species are clustered into homology groups (HGs) and aligned. For each HG, a gene tree is constructed and simplified by collapsing monophyletic branches composed exclusively of sequences from a single clade. This step minimizes noise from within-clade phylogenetic discordance and mitigates biases introduced by unequal sampling among clades (**Fig. 1a**). Recognizing that conventional homology clustering cannot reliably distinguish orthologs from paralogs, we implemented a tree-splitting algorithm to partition multi-copy HG trees into smaller, ortholog-like (OG-like) subtrees (**Fig. 1a**). Each OG-like tree is subsequently scanned to identify nodes exhibiting topological incongruence with the reference species phylogeny, as potential RI signals (**Fig. 1a**).

Besides introgression, ILS plays an essential role in shaping reticulate evolution patterns (Bjornson et al., 2024; Li et al., 2024; Stull et al., 2023). To distinguish RI from ILS, we incorporated a modified Branch-Length Test (mBLT) algorithm (**Supplementary Fig. 3**). Under ILS without introgression, triplet topology should exhibit no significant branch length disparity, compared to their sister nodes in the species tree. In contrast, introgression is expected to have discordant distribution, with the introgressed lineage nested within an otherwise divergent clade. Furthermore, to minimize false positives, we filtered phylogenetically ambiguous signals, including nodes with low bootstrap support, Bayes posterior probability or evidence of long branching attraction (**Supplementary Fig. 2**). For each high-confidence RI event, gene copies with incongruent phylogenetic placement are designated as acceptors, while their closest non-incongruent homologs are defined as donors.

To evaluate the performance of RIFinder, we simulated distinct RI levels across divergent clades (**Methods**). Across gene trees with varying RI frequencies, RIFinder consistently achieved high precision (85.7% to 98.6%; **Fig. 1b**). Particularly, at the 0.1% simulated transfer rate, it achieved a recall rate of 89.0% (**Fig. 1b**). In Poaceae, RI events were quantified to range from 0.03% to 0.3% of all genes, respectively (See next; **Supplementary Fig. 4**). At this biologically relevant transfer frequency (0.04% and 0.1%), RIFinder demonstrated robust performance, achieving high accuracy (95.9% to 98.2%, 93.5% to 96.3%) and high recall (85.0% to 87.6%, 83.5% to 84.0%) across the three scenarios (**Fig. 1b**). These results indicated that the simulation framework reflected biologically realistic introgression dynamics, thereby reinforcing confidence in the applicability of RIFinder to real-world genomes. Collectively, RIFinder offers a well-balanced solution for detecting RI, particularly in large and complex phylogenomic datasets.

## Widespread RI events across the grass genomes

We applied RIFinder to systematically identify RI events between PACMAD and BOP lineages in the grass family, by integrating high-quality genomes from 78 Poaceae species and outgroup species pineapple (*Ananas comosus*; **Supplementary Table 1**). Polyploid genomes were split into haploid subgenomes, resulting in a total of 122 haploid genomes, comprising 11 from CHL, 37 from PAN, 32 from BAM, 12 from ORY, and 29 from POO (**Fig. 2a**). The basal Poaceae species *Pharus latifolius* served as an additional outgroup (Ma et al., 2021). A final total of 5.15 million protein-coding transcripts were analyzed to detect DNA transfers between the PACMAD and BOP lineages. A total of 622 candidate RI events were identified across various divergence nodes between the two lineages, including 286, 64 and 272 events shared by singleton, doubleton and multiple species, respectively (**Fig. 2a**; **Supplementary Table 2**). RIFinder successfully recalled several previously characterized events with strong RI signals, such as the *Bx* cluster in Triticeae (Wu et al., 2022b), MABGC cluster in *Oryza* (Wu et al., 2022a) and a stress response-related DNA segment in barley (*Hordeum*) transferred from *Panicum* (Mahelka et al., 2021).

RI events have occurred recurrently along the diversification of grass species, with the PACMAD-to-BOP transfers exceeding the reverse (**Fig. 2a**). Intriguingly, a significant correlation between the number of acceptor and donor events per haploid genome was observed ($P = 5.85e-10$, two-tailed *t*-test), although the patterns varied by subfamilies (**Fig. 2b**). In the BOP lineage, POO and ORY have acquired an average of 117.7 and 67.3 genes per haploid genome, respectively, significantly more than BAM (**Supplementary Fig. 5**). In PACMAD, CHL species accepted fewer RI genes from BOP than those for PAN species. PAN contributed the highest number of transferred genes (~73.7 per haploid genome) relative to other subfamilies (**Supplementary Fig. 4a**). Interestingly, BAM exhibited the lowest numbers of both donor and acceptor gene counts among subfamilies, potentially due to its distinct life cycle (**Fig. 2a**). BAM grasses typically exhibit prolonged vegetative growth, followed by a synchronized, monocarpic flowering event, which limits temporal overlap with other flowering species and thus may reduce opportunities for inter-lineage gene transfer. Notably, a certain number of Poaceae species are primarily polyploidy, exhibiting subgenome dominance in gene expression and selection, with BAM as a representative example (Ma et al., 2024). In contrast to other subfamilies with no significant subgenome preference for RI-derived gene retention, BAM exhibited a striking subgenome bias in retention ratios ($p = 2.08e-10$, two-tailed *t*-test; **Supplementary Fig. 5c**).

In POO, Triticeae tribe (e.g. barley, rye and wheat) harbored significantly more acceptor genes than sister tribe Poeae (e.g. alkaligrass and oat), likely due to a Triticeae-specific RI pulse after the divergence between the two core Pooideae tribes (**Fig. 2a**). Molecular dating of each RI event in wheat (Triticeae) and oat (Poeae) revealed the recent RI pulse around 30 mya specific in Triticeae species, consistent with the divergence time between Triticeae and Poeae (~35.7 mya, Zhang et al., 2022) (**Fig. 2c**). For example, *Bx6* genes encode a 2-oxoglutarate-dependent dioxygenase (2-ODD) in the DIMBOA biosynthesis pathway for biotic defense (Jonczyk et al., 2008). Previous reports indicated that *Bx6* genes in POO species arose from the duplication of native homologs and were transferred from PAN (Wu et al., 2022b). Expanded sampling here revealed a novel *Bx6* homologous monoclade, termed *Bx6-like*, in Triticeae, phylogenetically nested within the PAN homologs (**Fig. 2d**; **Supplementary Fig. 6a**). Divergence time estimated between *Bx6* and *Bx6-like* clades with corresponding RI donor PAN homologs indicated that the *Bx6-like* RI event occurred more recently than the original *Bx6* event (**Fig. 2c**). The restricted presence of *Bx6-like* genes in Triticeae supported the notion of a lineage-specific RI. Topology tests confirmed that the phylogeny of duplicated *Bx6* genes was highly robust and supported the introgression from PAN to Triticeae (**Supplementary Fig. 6b**). Expression of *Bx6-like* gene TraesCS2B01G076400 in the hexaploid wheat (*Triticum aestivum*) was induced by multiple stresses, comparable to the native copy TraesCS1B01G103700 in the B subgenome and other *Bx6* copies (TraesCS2B01G066000 and TraesCS2A01G051700; **Supplementary Fig. 7**).

## Characteristics and functional implication of RI genes

The encoded proteins of RI genes demonstrated pronounced involvement in stress response processes, although the functional enrichment varied across subfamilies **(Fig. 3a-d)**. In wheat, we observed significant enrichment of wax ester synthase domains (PF03007, $p$ = 4.76e-45; PF06974, $p$ = 1.15e-58; Fisher's exact test), critical for the wax formation of leaf surfaces, an adaptive trait enhancing tolerance to both biotic and abiotic stress (Aharoni et al., 2004; Bernard and Joubès, 2013). In rice, domains such as lipoxygenase and NB-ARC, known for pathogen and parasitic fungus defense, were significantly over-represented (**Fig. 3a**). Additionally, the enrichment of terpene synthase domains suggested a potential role in secondary metabolite biosynthesis, like momilactone, from RI genes. In maize, aconitase-related RI genes, implicated in stress responses, were significantly enriched (Meng et al., 2022; Pascual et al., 2021; Selinski et al., 2018). Collectively, RI-derived genes frequently encode proteins involved in signaling and stress responses, potentially enabling dynamic modulation of gene expression and immune strategies in acceptor species.

As foreign copies of native genes, RI-derived genes may share evolutionary fates with gene duplicates, undergoing loss or retention through mechanisms such as compensatory drift or sub-functionalization (Birchler & Yang, 2022). We profiled the transcriptomic dynamics for the RI accepted genes in rice and wheat (**Supplementary Tables 3 and 4**). Among the 54 RI foreign genes in rice, 40 were expressed in at least one tissue (FPKM > 1). Notably, *GF14a* exhibited constitutive expression across all tissues examined (**Fig. 3e**). This gene was reported to repress the activity of *OsMYBS2* to protect against photodamage and photoinhibition under excess light conditions (Fu et al., 2021). Additionally, *GF14a* resides within a selective sweep region, implying its role in domestication (**Fig. 3e**). Under stress treatments, 40 genes displayed differential expression (**Fig. 3e**). *Bph14* was induced by flood treatment but downregulated during brown planthopper infestation (**Fig. 3e**). Under phosphate starvation and salt stress, two homologs (LOC_Os01g37000 and LOC_Os06g16640) exhibited divergent expression patterns compared to the native ortholog (LOC_Os07g38590), suggesting their functional specialization in stress adaptation (**Supplementary Fig. 8**).

In contrast, gene expression patterns in wheat were influenced by polyploidy and subgenome asymmetry (Ramírez-González et al., 2018; Yang et al., 2021). Among the 191 foreign genes, 112 (58.6%) were expressed, distributed across the A ($n$ = 31), B ($n$ = 42), and D ($n$ = 39) subgenomes, with 28 genes exhibiting tissue-specific expression (**Fig. 3f**). Additionally, 179 RI genes derived from polyploidization or duplication were retained across subgenomes, in which 145 were preserved in all three subgenomes. Notably, 52 gene copies showed no detectable expression under any treatment, likely due to dosage compensation or transcriptional silencing (**Fig. 3f**). This indicates that, compared to rice, the polyploid structure of wheat genome has introduced greater complexity in the RI-derived gene regulation, characterized by subgenome-specific retention and divergent functional trajectories among homeologs.

## A Triticeae-derived segment in Cleistogenes songorica associated to drought tolerance

The tetraploid CHL species *Cleistogenes songorica* (2n = 4× = 40, abbreviated as Cson) is a vital perennial forage grass and ecologically important C4 species, thriving in temperate saline, semi-arid, and desert regions of Central Asia, with exceptional drought tolerance (Yan et al., 2019; Zhang et al., 2021; **Fig. 4a**). A 30-Kbp segment on chromosome 6 in the A subgenome (Cson-A), comprising five protein-coding genes CsA601273 (*RIG1*), CsA601274 (*RIG2*), CsA601275 (*RIG3*), CsA601276 (*RIG4*), and CsA601277 (*RIG5*), was identified as a RI fragment (**Fig. 4b**). Among these genes, *RIG1* and *RIG5* are homologous to *OsMATE2* and *OsPSKR15*, both implicated in abiotic stress responses (**Supplementary Table 5**) (Das et al., 2018; Nagar et al., 2022). Phylogenetic analyses revealed that these RI-acquired genes in Cson-A clustered within the POO clade and were sister to *Achnatherum splendens* (Aspl; 2n = 2× = 48), a Triticeae species dominant in alkaline grasslands (**Figs. 4a-b**). Notably, donor genes in Aspl resided within a ~20-Kbp region with identical gene order and orientation, suggesting a single RI event (**Fig. 4c**). Local sequence alignment demonstrated high similarity between the two gene clusters, particularly in the protein-coding regions (**Supplementary Fig. 9**).

Notably, this RI segment was exclusively present in Cson-A but absent from the B subgenome (Cson-B) and other CHL genomes, implying recent lineage-specific acquisition (**Figs. 4b-c**; **Supplementary Fig. 10**). Technical artifacts were ruled out by uniform read coverages across RI donor and acceptor segments and absence of clipping signals at their boundaries, when raw sequencing reads from Cson and Aspl were remapped to their assemblies, respectively (**Supplementary Fig. 9**). Additionally, a homologous region containing CsB800052 (*HRIG2*), CsB800053 (*HRIG3*), CsB800054 (*HRIG5*), was observed in Cson-B. Phylogenetic analysis confirmed that these B-subgenome genes as vertically inherited copies, distinct from the horizontally acquired cluster in Cson-A (**Fig. 4b**). Protein sequence domain alignments further supported the evolutionary affinity between the RI-acquired genes in Cson-A and RI donor genes in Aspl (**Supplementary Fig. 11**). Topology tests using nucleotide alignments revealed that introgression models were significantly better supported than non-RI hypotheses (**Supplementary Fig. 12**). Collectively, these results strongly indicated that this gene cluster in Cson-A originated from Aspl or its close relatives via RI. Interestingly, the RI donor cluster in Aspl has a segmental duplicate (SegDup) on chromosome 13, absent in other POO species (e.g. barley and wheat; **Figs. 4b-c**). Non-RI homologs of this cluster exhibited conserved genomic orientation across grass species, implying their presence predated grass divergence (**Fig. 4c**). Sequence divergence analyses revealed that the Cson-A cluster derived from the Aspl RI donor rather than its SegDup (**Fig. 4d**). Expression profiling of RI-acquired genes and their native homologs under stress revealed no co-expression (**Supplementary Fig. 13**). Under drought stress, *RIG2* and *RIG3* were significantly down-regulated, whereas *RIG5* was up-regulated (**Fig. 4e**). Among native homologs, only *HRIG2* showed stress response. Similar expression patterns were observed under high-temperature stress (**Fig. 4e**). These findings suggested that the RI cluster in Cson-A from Aspl has undergone functional divergence from its native counterparts, potentially contributing to its exceptional drought tolerance.

Geographically, Aspl and Cson share a sympatric distribution in East Asian saline-alkaline grasslands, whereas Cson's phylogenetically closest species, *Zoysia japonica*, is predominantly restricted to the coast region of East Asia and coastal Australia, and *Oropetium thomaeum* is endemic to Africa (**Fig. 4f**; **Supplementary Fig. 14**). Climatic niche differentiation analyses indicated substantial overlap between Aspl and Cson (Warren's $I = 0.760$ and the Schoener's $D = 0.572$), relative to the null hypothesis of niche equivalency (**Supplementary Fig. 15a**), whereas Aspl/*Z. japonica* (Warren's $I = 0.0388$, Schoener's $D = 0.1028$, $p < 0.001$) and Cson/*Z. japonica* (Warren's $I = 0.0157$, Schoener's $D = 0.0714$, $p < 0.001$) displayed strong partitioning (**Supplementary Fig. 15b**). Species density data along two climatic gradients further showed the niche space of Cson entirely nested within that of Aspl (**Fig. 4g**). These findings implied the potential relationship between geographic niche proximity and the preserved cross-subfamily gene flow between Aspl and Cson, from broad-range species to narrow-range species.

## Evolutionary trajectory of gramine biosynthetic gene cluster among grass

Gramine, an allelopathic indole alkaloid, exhibits broad-spectrum defensive activities against aphids, fungi, viruses, and other plants, which has been reported to be biosynthesized in certain grass species (e.g. *Hordeum*) (Brown and Trigo, 1995; Corcuera, 1984; Leete and Minich, 1977; Lu et al., 2019). Two core biosynthetic genes *AMIS*/*CYP76M57*, encoding a cytochrome P450 monooxygenase, and *NMT*, encoding a N-methyltransferase, have been demonstrated to mediate gramine production in wild barley *H. vulgare* subsp. *spontaneum* and specific cultivated barleys (*Hordeum vulgare*). These genes are physically adjacent in the genome (**Fig. 5a**) (Dias et al., 2024; Ishikawa et al., 2024), with *AMIS* encoding enzyme oxidizing tryptophan into aminomethylindole (AMI), and *NMT* enzyme methylating AMI twice via the intermediate N-methylaminomethylindole to yield gramine (**Fig. 5a**). Importantly, genes within the gramine biosynthetic gene cluster (GBGC) exhibited RI signals across grass subfamilies (**Fig. 5b**).

Phylogenetic analysis of *AMIS* homologs defined three clades: a positionally conserved clade (Clade 1, where the genes were phylogenetically clustered and colocalized in syntenic blocks across grass species), an ORY-specific clade (Clade 2, encompassing four core P450 genes from *CYP76M5 to CYP76M8* involved in phytocassane biosynthesis on chromosome 2), and a GBGC-associated clade (Clade 3). Phylogeny discordance was primarily observed in Clade 3, although a small cluster of *Oryza* genes nested within PACMAD in Clade 1 (**Fig. 5b**). Basal members in Clade 1 (Aluo-C_Alu08Cg07060) and Clade 3 (Aluo-C_Alu08Cg07070) were tandem duplicates in the *Ampelocalamus luodianensis* genome from BAM, indicating their common origination (**Fig. 5b**). BAM homologs from Clade 1 were found across multiple bamboo subgenome types, including the ancestral type H in herbaceous bamboos, whereas Clade 3 homologs were confined to type C genomes, the sole subgenome shared by all polyploid bamboo species, and entirely absent from other subgenomes, suggesting subgenome-specific retention and selection-driven specialization (Ma et al., 2024). Topology of Clade 3 was statistically supported and incongruent from the species tree (**Supplementary Fig. 16**), with homologs from *Eragrostis* (CHL), *Cenchrus* (PAN), and POO species forming a monoclade sister to the *Oryza* lineage. Within POO, *AMIS* from wild barley (*H. vulgare* subsp. *spontaneum*) occupied a basal position, while its two duplicates clustered with other POO homologs, particularly in Triticeae species, which exhibited extensive tandem duplication (**Fig. 5b**).

Phylogenetic topology of *NMT* homologs mirrored that of *AMIS* genes with significant RI signals, where CHL, PAN and POO copies were nested within ORY when BAM species were set as outgroups (**Supplementary Figs. 17-18**). Unlike *AMIS*, *NMT* homologs were largely absent from POO, except in *Hordeum*. Taken together, besides *Hordeum*, gene clustering of *AMIS* and *NMT* was identified in both ORY (e.g. *O. sativa*, *Z. latifolia*) and PAN (e.g. *Cenchrus*; **Fig. 5b**) lineages. Synteny comparisons revealed identical gene orientations, supporting the hypothesis of co-introgression scenario for the two genes, despite their physical separation by several kilobases (**Fig. 5c**). By comprehensively integrate the phylogeny and synteny evidence of the two genes, and subfamily lineage divergence time, we infer that the gathering of *AMIS* and *NMT* emerged firstly in the ancestral lineage of BOP, showing an ILS pattern when comparing the gene topology to species phylogeny (Grass Phylogeny Working Group III, 2024). Subsequently, a remote introgression event occurred into the common ancestor of PAN and CHL, whose divergence time was much younger than the radiation divergence among BOP subfamilies (**Fig. 5c**).

In rice, two additional *NMT* homologs, *OsNMTa* (LOC_Os12g10140) and *OsNMTb* (LOC_Os12g10170), were present near the *AMIS-NMT* cluster (< 200 Kb), derived from duplication of *OsNMT* (LOC_Os12g09770) and subsequent fission (**Supplementary Figs. 19-20**). RNA-seq data showed up-regulation of these genes under blast infection and nitrogen starvation (**Supplementary Fig. 21**). Co-expression network analysis revealed that *OsNMTb* co-expressed with MABGC (**Supplementary Fig. 22a**). However, similar to the presence-and-absence variation polymorphism of GBGC in barley (**Fig. 5c**; **Supplementary Fig. 23**), this genomic region was almost exclusively observed in *O. sativa* ssp. *japonica*/*geng* and its wild ancestor group Or-3 of *O. rufipogon*, as shown from the rice pangenome graph (**Fig. 5e**; **Supplementary Fig. 24**). Moreover, this region was under artificial selection in the domestication of *O. sativa* ssp. *japonica*/*geng* (**Supplementary Fig. 22b**). In wheat, two tandem *AMIS* homologs exhibited strong induction by *Fusarium* infection, and co-expressed with neighboring gene *TaELL1*, encoding tryptamine 5-hydroxylase, indicating potential involvement in the 5-hydroxytryptamine biosynthesis pathway, contributing to pathogen resistance (**Fig. 5g**). Similarly, extensive structural variations were observed in wheat pan-genomes, reflecting balancing selection on such immunity related locus (**Supplementary Fig. 25**).

To assess the ability of gramine biosynthesis in grass species with GBGCs, we firstly quantified the expression of GBGC genes in the fresh leaves of *Z. latifolia*, *C. macrourus, O. sativa* by methyl jasmonate (MeJA) treatment, with barley serving as a positive control. Significant up-regulation was observed for barley GBGC genes, but no detectable expression in *O. sativa* or *Z. latifolia*, even under MeJA treatment (**Supplementary Fig. 26**). In *C. macrourus*, the GBGC genes were transcriptionally active under MeJA-treated conditions, though at low expression levels (**Supplementary Fig. 26**). Correspondingly, using extracted ion chromatograms (EICs), gramine-target metabolic profiling detected trace levels of gramine in leaves at control and 4-hour post-treatment (**Fig. 5h**). In contrast to the distinct peak in barley highly aligning with the standard, three left-sided gramine-related ion peaks (designated as unknown 1, 2, and 3) exhibited response trends in abundance under MeJA stress (**Fig. 5h**). These compounds may represent functionally gramine derivatives or isoforms that respond to biotic stress in *C. macrourus*, rather than gramine observed in barley, which requires further investigation.

# Discussion

Beyond the well-documented instances of cross-kingdom HGT and intra-genus introgression, RI represents an under-explored mechanism of adaptive gene acquisition (Whitney et al., 2010; Wu et al., 2024; Yang et al., 2019). Here, we developed a phylogeny-based workflow optimized for RI detection. Compared to previously sporadic reports (e.g. momilactone A biosynthetic cluster in rice, stress-resistance gene transfers from *Panicum* to wild barley) (Mahelka et al., 2021; Wu et al., 2022a), we detected a total of 622 RI events between BOP and PACMAD lineages using RIFinder, providing novel insights into adaptive evolution of grass species. To highlight the functional significance of RI, two representative examples were comprehensively investigated: the drought-related gene cluster in Cson (CHL) likely derived from Aspl relatives (POO), and gramine-associated genes in *Hordeum* (POO) and *Cenrus* (PAN) from ORY.

It should be noted that distinguishing true RI from ILS still challenges the accuracy of RI detection, although we have implemented a modified branch-length test to exclude ILS-driven topological incongruence. Additionally, the current framework is constrained by limited phylogenetic resolution. While selective retention may confine RI signals to regions near functional genes, future improvements will enhance resolution and detect a broader spectrum of RI events, by integrating a *k*-mer-based or pangenome-alignment framework. Given our conservative emphasis on accuracy over sensitivity, and considering taxonomic under sampling and challenges in distant homology detection, our reported RI events likely represent a lower-bound estimate.

Taking grass species as an example, uneven distribution of RI events across subfamilies likely reflects biological and ecological implication, including life history (including reproductive cycle), and geographical/niche distribution. In previous studies, exposure of reproductive tissues during early development facilitates gene acquisition in plants (Ma et al., 2022; Guan et al., 2018; Huang et al., 2013). Our findings point to distribution range and reproductive timing as putatively critical variables influencing RI propensity in grasses. For instance, a long reproduction cycle in bamboo might lead to reduced opportunities for gene exchange (Yeasmin et al., 2015). Additionally, we observed asymmetry in acceptor and donor gene numbers between BOP and PACMAD lineages. Species from PACMAD act as donors more frequently, whereas BOP lineages tend to serve as acceptors in RI processes. This imbalance may reflect differences in ecological and geographic distributions between the two clades. For instance, POO species are primarily temperate, ORY species are tropical/subtropical, and BAM species are restricted to South America, Southern Africa and Southeast Asia. In contrast, PAN and CHL span a broader latitudinal and ecological range, from the Arctic Circle to the Cape of Good Hope, resulting in less likely to fix (**Supplementary Fig. 27**).

Regarding genetic mechanisms underlying RI, two evolutionary scenarios are plausible: introgression between ancestral lineages of extant species, or hybridization between evolutionarily divergent taxa that have overcome reproductive barriers (Lan et al., 2023; Seehausen et al., 2008). Ancestral introgression likely predates the establishment of reproductive isolation, aligning with the concept of a ‘speciation continuum’ (Kulmuni et al., 2020; Nosil et al., 2009; Seehausen et al., 2008), particularly in clades undergoing radiation, where incomplete reproductive barriers may fail in preventing genetic exchanges (Marques et al., 2019; Pontarp et al., 2024). Nevertheless, the detection of such ancient events is complicated by ILS, which may obscure introgression signals (Stull et al., 2023). Empirical evidence from diverse taxa supports the feasibility of distant hybridization as a pathway of RI. In primates, a recent study identified Y chromosome introgression in *Cercopithecini* monkeys, estimated at ~8 mya (Jensen et al., 2024). In plants, ancient introgressions have been recurrently detected in bamboo genomes (Guo et al., 2019). Recent experimental work demonstrates that reproductive barriers can be overcome between distantly related species. In *Arabidopsis*, hybrid embryos were successfully generated between species diverged for >14 mya via pollen manipulation (Lan et al., 2023). In *Brassica*, the removal of reactive oxygen in the stigma alleviated the intergeneric reproductive barrier between *Arabidopsis* and *Brassica* (Huang et al., 2023). These findings across diverse taxa collectively indicate the possibility of RI via ancestral gene flow and naturally facilitated hybridization across deeply divergent lineages.

# Methods

## RIFinder workflow

RIFinder is a Python-based tool developed to detect RI events from large-scale genomic datasets. It is designed to identify phylogenetic discordance arising during species divergence, which may signal introgression between distantly related taxa. It integrates modules of gene tree inference, signal quantification, and statistical testing (**Supplementary Fig. 2**). The workflow begins with reconstructing gene trees for homologous gene groups, incorporating stringent quality control to remove unreliable phylogenies (e.g. those with abnormal branch lengths or poorly supported nodes). These curated gene trees are compared to species tree to identify topological discrepancies. RIFinder quantifies RI values for each gene tree using a topology-based scoring formula. To ensure specificity, a statistical significance module tests whether observed signals exceed expectations under neutral ILS scenario. By integrating these steps into a unified and automated framework, RIFinder leverages multi-genome data to accurately capture reticulate evolutionary processes that are frequently overlooked in conventional tree-based models.

## MCL clustering and gene phylogeny

Pairwise protein sequence alignments were performed across all genome assemblies using DIAMOND (v2.0.2) with an E-value threshold of 1e-5 and a minimum query coverage of 50% (Buchfink et al., 2021). Based on these similarity scores, protein sequences were clustered into homologous groups (HGs) using the Markov Clustering Algorithm (MCL) with an inflation parameter of 2. For the HGs with sufficient taxonomic representation, multiple sequence alignments were conducted using MAFFT (Rozewicki et al., 2019), and poorly aligned regions were pruned using TrimAl (Capella-Gutiérrez et al., 2009). Substitution models were selected for each alignment using ModelFinder (Kalyaanamoorthy et al., 2017), and gene trees were inferred under a maximum-likelihood framework using IQ-TREE (v2.0) with 3,000 ultra-fast bootstrap replicates (Minh et al., 2020).

## Compressing and splitting gene trees

Upon initializing RIFinder, the pipeline performs quality control of gene trees by removing extreme outliers, characterized by long branch lengths. For each gene tree, the distances from each leaf to its three immediate ancestral nodes are calculated. Any leaf with a distance exceeding the 99th percentile of distance distribution within the tree is flagged and removed. After outlier removal, the algorithm traverses each node in the gene tree. For monophyletic groups from a single defined clade, a node-collapsing operation is applied to simplify the tree by reducing intra-clade topological complexity (**Fig. 1a**). Paralogous genes, especially those resulting from ancient or lineage-specific duplication, are often included in gene family trees, which complicate orthologs inference and phylogenetic reconstruction. As similarity-based clustering methods tend to group orthologs and paralogs, large gene families may be artificially inflated. To address this, RIFinder implements a tree decomposition algorithm to split large gene trees into smaller, orthologous group-like (OG-like) clusters.

The principle of Tree splitting algorithm is described using the following pseudocode:

*FOR a rooted gene family phylogeny:*

*OGlike ← [];*

*> forall leaf of gtree do*

*>> cur node ← leaf;*

*>> while node comprising clade ≤ whole clade do*

*>>> node ← node.up;*

*>> end*

*>> AddToList(OGlike, node);*

*> end*

*return Oglike*

*ENDFOR*

RIFinder iterates through the leaves of each gene tree to assess whether the current subtree contains taxonomically diverse representation. Specifically, it checks if the number of unique clade identifiers in the subtree meets a user-defined threshold (default: $\sqrt{n}$, where $n$ is the total number of taxa in the tree). If a current subtree fails the criteria, the algorithm moves upward to the parent node and re-evaluates the taxonomic diversity. Once a subtree satisfies the inclusion criteria, it is extracted for downstream RI analysis; otherwise, it is excluded from further consideration. This sequential process of outlier filtering, node collapsing and tree decomposition ensures that RIFinder operates on high-confidence, taxonomically representative gene subtrees, thereby enhancing the resolution and reliability of RI signal detection across complex multi-copy gene families.

## RI scoring

Following pre-processing and filtering, gene phylogenies are screened for topology indicative of RI (**Fig. 1a**). Species phylogeny is divided into two discrete taxonomic clades, representing the major groups for investigation. Species are classified as either congruent (i.e. belonging to the same monophyletic clade) or discordant (i.e. falling outside the expected monophyletic group) per gene tree. This classification aligns with the taxonomic level defined for RI. RI scores are computed for leaves in the gene tree based on taxonomic identity of sister branches, the number of discordant taxa, and their distribution across the tree. The threshold is determined using Cayley's formula: $n^{(n-2)}$, where $n$ denotes the number of taxa in a given sub-clade. This reflects the number of distinct unrooted tree topologies that can be formed from $n$ tips. Any leaf whose RI score exceeds that is labeled as a candidate RI event (**Fig. 1a**). To calculate RI scores within each OG-like subtree, the algorithm starts at a target leaf that meets the minimum taxa threshold, iteratively. At each internal node, the algorithm examines the newly incorporated descendant branches to identify taxa that deviate from the expected monophyletic group. Scoring begins upon detection of the first discordant taxon and continues until a taxon from the original clade reappears, at which scoring stops. Three distinct scoring scenarios are implemented. If all newly added leaves belong to the same discordant clade, their total number is added directly to the node's RI score. If discordant taxa originate from multiple distinct clades, a fractional score is calculated as the ratio of discordant taxa to the total number of newly added taxa. If not, all expected discordant clades are represented, the tree is excluded from downstream analysis to avoid potential bias from rooting artifacts or incomplete sampling. The full logic of the RI scoring procedure is outlined in pseudocode format as follows.

```
Input: A gene tree topology
CT ← cladei;
CN ← cladej;
scoreL ← [];

> forall leaves do
>> score ← 0;
>> clade ← leaves;
>> dclade ← !leaves;

>> if taxon < CT then
>>> Continue
>> else
>>> node ← leaves;
>>> while node parent exist do
>>>> node ← node.up;
>>>> newTaxn ← getNewleaf(node);
>>>> allTaxn ← getLeaf(node);
>>>> if dclade in allTaxn then
>>>>> score ← score + weight;
>>>> if clade in newTaxn then
>>>>> Break
>>>> end
>>> end
>> end
>> AddToList(scoreL, score);
> end
end
Return scoreL.
```

To ensure result robustness, only nodes with bootstrap over 70 (default) are retained. In gene family trees with multiple plausible outgroup placements, the scoring process is repeated across alternative roots, and RI scores are aggregated to mitigate rooting bias.

## Modified Branch-Length testing

After identifying candidate RI events, each case was further evaluated to rule out alternative sources of phylogenetic incongruence, with a primary focus on ILS. To distinguish RI from ILS, we developed a modified branch-length testing (mBLT), inspired by the framework proposed by Suvorov et al. (**Supplementary Fig. 3**). The mBLT estimates relative coalescence times by analyzing branch lengths in gene trees, measured as substitutions per site. Under a RI scenario, the most recent common ancestor (MRCA) between the donor and recipient taxa is expected to exhibit shorter branch lengths than expected under ILS alone. In contrast, ILS typically leads to older coalescence time and does not generate consistent directional differences across alternative discordant topologies. This forms the null hypothesis of mBLT: in the absence of introgression, coalescence time should not differ significantly between topologies. For each candidate RI event, three clades, acceptor, donor, and sister, are defined as a triplet. Pairwise distances (denoted as $d$) are computed based on external branch lengths between leaves from the recipient clade and those from the donor or sister clade, respectively. Specifically, for each leaf in the recipient clade, distances to all donor leaves $d_{(Donor,\ Accepted)}$ and all sister clade leaves $d_{(Donor,\ Sister)}$ are measured. The distributions of $d_{(Donor,\ Accepted)}$ and $d_{(Donor,\ Sister)}$ are then compared using a two-tailed independent $t$-test. Under pure ILS, the expected relationship is $d_{(Donor,\ Accepted)} \approx d_{(Donor,\ Sister)}$, whereas $d_{(Donor,\ Accepted)} < d_{(Donor,\ Sister)}$, indicates a possible introgression event consistent with the discordant topology. Conceptually, the mBLT aligns with previously described methods such as the original BLT (Suvorov et al., 2022) and the $D_3$ statistic (Hahn and Hibbins, 2019), leveraging topological information across loci to detect introgression.

For genome-wide application, we apply mBLT to all homologous gene trees in the pangenome, and $p$ values are adjusted using the Benjamini–Hochberg procedure with a false discovery rate (FDR) threshold of 0.05. It is important to note that the mBLT has reduced sensitivity when introgression occurs at similar rates into both the recipient and sister clades, or when gene flow took place in a deep ancestral lineage. Once a candidate RI event is confirmed, associated metadata, including recipient and donor clades, discordant topologies, and relevant annotations, are recorded (**Supplementary Table 2**). Donor and accepted copies identities are assigned based on their respective last common ancestors in the species tree. Additionally, in cases where a subset of leaves within a discordant clade fail to meet the criteria but are phylogenetically nested within the target clade, the corresponding RI event is annotated as recent introgression.

## Topology testing

For individual phylogenies of particular interest, analyses are repeated following the same procedures as described above, but with more rigorous alignment settings in MAFFT (i.e., additional iterations) and limited manual curation, including the removal of long-branching or poorly aligned taxa. Substitution models are reselected using ModelFinder, and maximum-likelihood phylogenies are reconstructed using IQ-TREE, with ultra-fast bootstrap support computed over 3,000 replicates (Minh et al., 2020). Resulting trees are visualized and annotated using iTOL (v5; Letunic and Bork, 2021). To test the robustness of alternative topologies, we apply three tree topology comparison methods implemented in IQ-TREE: the RELL approximation, the Shimodaira-Hasegawa (SH) test, and the Approximately Unbiased (AU) test, each based on 10,000 bootstrap replicates (Minh et al., 2020). AU and SH tests yielded $p$ values, and alternative topologies are rejected if $p < 0.05$. The RELL approximation is used to compute posterior weights for each tested topology (Kishino et al., 1990). To minimize the confounding effects of alignment gaps, TrimAl is used to remove poorly aligned regions, and the resulting conserved alignments are used for topology testing (Capella-Gutiérrez et al., 2009). In addition, phylogenetic trees are constructed based on nucleotide coding sequences (CDS), as well as separately for the first and second codon positions (codon12) and third codon positions (codon3), following the same maximum-likelihood framework.

## Simulation

Benchmark datasets were generated under three distinct topologies using GenPhyloData in JPrIME v0.3.7 (Sjöstrand et al., 2014). These topological structures represent a variety of configurations under a rooted bifurcating framework, ranging from simple (2 *versus* 2) to complex (3 *versus* 4) scenarios, with each subgeneric clade comprising multiple species forming monophyletic groups. The bifurcating species trees for the three topological relationships were generated by HostTreeGen using the parameters of '-timespan 10, -birth rate 0.3 and -death rate 0.1', with each subgeneric clade containing more than 4 species. The birth-death process was employed to generate species trees, with extinct lineages systematically pruned from the final phylogenies. These three resultant species trees, containing 33, 45, and 59 extant taxa respectively, each including an additional outgroup lineage, served as backbone trees for subsequent gene tree simulations. To maintain experimental consistency across simulations, all gene trees were generated by GuestTreeGen under identical parameters: duplication rate = 0.2 and loss rate = 0.1 per million years, with leaf node counts constrained to 1-4 times the species number. An additional constraint limited each species to ≤10 gene leaf nodes in any simulated gene tree. To simulate diverse biological contexts of RI, we generated three distinct datasets corresponding to transfer rates of 0.1%, 0.4%, 0.7%, 1%, and 2% per lineage-million years. A strict filtering protocol was implemented to curate RI: 1, Outgroup-associated RI events were removed to avoid ancestral lineage interference; 2, Intra-subclade transfers were excluded to focus on biologically meaningful transfer. Subsequently, branch lengths were temporally calibrated using the BranchRelaxer module (Sjöstrand et al., 2014), which employs a LogNormal-distributed relaxed molecular clock to rescale gene tree branches while preserving topological congruence with species trees. The simulation pipeline yielded 7,402 high-confidence gene trees containing RI events, with each transfer rate condition (0.1%/0.4%/0.7%/1%/2%) producing > 150 qualifying trees meeting all evolutionary criteria. The performance of RIFinder was quantified through rigorous comparison against introgression events reported in these datasets.

## Grass genomes

All genomic sequences and gene model annotations from 78 high-quality grass assemblies, and outgroup species *A. comosus* were sourced from Phytozome (https://phytozome-next.jgi.doe.gov), National Genomics Data Center (NGDC) (https://ngdc.cncb.ac.cn) and Ensemble Plants (https://plants.ensembl.org). These genomes included representatives from the grass basal group (*P. latifolius*), Oryzoideae (*Z. latifolia*, *L. perrieri* and *Oryza* spp.), Bambusoideae (*Dendrocalamus* spp., *Ampelocalamus luodianensis*, *Bonia amplexicaulis*, *Guadua angustifolia*, *Hsuehochloa calcarean*, *Melocanna baccifera*, *Olyra latifolia*, *Otatea glauca*, *Phyllostachys edulis*, *Raddia guianensis*, *Rhipidocladum racemiflorum*), Pooideae (*Avena* spp., *Alopecurus myosuroides*, *Achnatherum splendens*, *Aegilop tauschii*, *Brachypodium distachyon*, *B. stacei*, *D. villosum*, *H. vulgare*, *Lolium multiflorum*, *Puccinellia tenuiflora*, *Secale cereale*, *Triticum* spp., *T. elongatum*), Chloridoideae (*Eragrostis* spp., *E. indica*, *O. thomaeum*, *C. songorica*, *Z. japonica*), and Panicoideae (*Alloteropsis semialata*, *Cenchrus* spp., *Digitaria* spp., *Echinochloa* spp., *Zea mays*, *S. bicolor*, *Miscanthus sinensis*, *Dichanthelium oligosanthes*, *Pennisetum giganteum*, *Panicum* spp., *Setaria* spp., *Paspalum vaginatum*, *Saccharum spontaneum*). Polyploid genomes with chromosome-level assemblies were split into subgenomes (**Supplementary Table 1**). A final of 122 grass haploid genomes were used in analysis.

## Genomic synteny

Whole-genome protein sequences were compared pairwise among the 122 genomes using BLASTP. Best hit for each blast search was retained and the threshold was set with the e-value less than 1e-5 and the identity greater than 50%. According to the physical positions of the genes on each chromosome of each species, the genes or proteins were ordered. Then, gene-to-gene synteny analysis among grass species was conducted based on gene orders within each genome. Micro-synteny analysis was conducted with JCVI (Tang et al., 2024). Syntelog-based pangenome was constructed using SynPan (https://github.com/dongyawu/PangenomeEvolution) using 119 chromosome-level assemblies.

## Functional enrichment

All predicted protein-coding sequences were annotated by InterProScan (v.5.65) (Zdobnov et al., 2001) and Pfam domains were determined with default parameters, along with KEGG. Gene families were identified based on Pfam annotations. In enrichment analysis, all genes across grass assemblies were used as background. We tested the null hypothesis that the Pfam associated with RI reflects a random sampling of total protein families. To this end, each term was compared against the background by Fisher's exact test. Fold change was calculated based on the number of each Pfam in the subfamilies.

## Expression and selection of RI acceptor genes in rice and wheat

Uniformed expression data for rice and wheat from multiple tissues under various treatments, are adopted form the Plant Public RNA-seq Database (PPRD, https://plantrnadb.com/) (Yu et al., 2022). Specifically, we selected 618 libraries for rice and 476 libraries for wheat (**Supplementary Table 6**). For rice, we collected transcriptome data from 3 varieties, covering 11 different tissues and 18 distinct treatments (15 abiotic and 3 biotic. For wheat, the dataset included 11 different tissues and 13 distinct treatments (5 abiotic stress and 8 biotic stresses). Accepted genes with putative native copy in rice and wheat were extracted according to phylogeny and further checked to generate accepted-native pairs manually. To investigate the evolutionary significance of potential RI genes by artificial selection in domestication, we screened for overlap between candidate RI genes and previously reported selective sweep regions in rice and wheat (He et al., 2019; Jing et al., 2023).

## Expression of RI accepted genes in *C. songorica*

Available 124 RNA-seq libraries of *C. songorica* were collected from NCBI and subsequently mapped to the *C. songorica* genome sequence using Tophat2 (**Supplementary Table 6**) (Kim et al., 2013). Gene expression of each sample was quantified with featureCounts and FPKM was computed with a custom script (Liao et al., 2014). Differential expression analysis was performed using the pydeseq2 (Muzellec et al., 2023). FDRs were adjusted using the posterior probability. FDR $< 0.05$ and $|\log_2 \text{(Fold-change)}| \geq 2$ were set as the threshold for significantly differential expression. The block-wise co-expression network was constructed and the module was identified with R package WGCNA (v.1.66; Langfelder and Horvath, 2008). The number of modules was detected automatically, with the number of genes in a module limited between 20 and 5000 genes.

## Geographical distribution

Distribution ranges of all species involved in the study were obtained from the GBIF (https://www.gbif.org). For subfamilies distribution, we traversed the distribution range of each genus, and integrated them according to the phylogenetic classification. After removing locations with fewer than 10 occurrences, the final distribution maps were created. For the distribution ranges of *C. songorica*, *A. splendens, O. thomaeum* and *Z. japonica*, we removed locations with only one appearance and visualized the distribution ranges using heatmaps.

## Ecological niche comparison

The initial dataset included 514, 55, and 940 records of *A. splendens*, C. songorica, and *Z. japonica*, respectively. To reduce sampling bias in niche modeling, occurrences with pair-distance < 5 km were excluded from the raw dataset using the thin function in the R package spThin (v. 0.2.0) (Aiello-Lammens et al., 2015). The data trimming process resulted in 681 records (199 of *A. splendens*, 35 of *C. songorica*, and 445 of *Z. japonica*). Subsequently, we retrieved near-present (1970-2000) climate layers of 19 bioclimatic variables (BIO1-BIO19) at 2.5 min spatial resolution from the WorldClim database (Fick & Hijmans, 2017). Five uncorrelated bioclimatic variables (|Pearson's $r$| < 0.7; BIO1, BIO2, BIO3, BIO8, and BIO15) were selected to characterize the climatic niche volumes. Based on Warren's $I$ and Schoener's $D$ statistics, we estimated pairwise niche equivalency among the three species using the *niche.equivalency.test* function (with 100 replicates) in the R package ecospat (v. 4.1.2) (Warren et al., 2008; Schoener, 1968; Di Cola et al., 2017). To provide an intuitive visualization of the climatic niche space occupied by the three species, we partitioned the climatic niches into two gradients using principal component analysis and plotted the species density along these gradients with the *plot.niche.dyn* function in the R package ecospat.

## Pangenome graph and local sequence alignment

We used 70 rice genome assemblies to construct a local pangenome graph of the GBGC locus in rice (Xie et al., 2024). Genomic coordinates of GBGC homologs across these assemblies were identified via BLAST searches using the T2T-NIP sequence as query (Shang et al., 2024). For each best hit, a 1-Mb interval (±500-Kb flanking region) was extracted and used to build the local graph with Minigraph (v0.21-r606, Li et al., 2020). The graph structure was visualized using Bandage (v0.8.1, Wick et al., 2015). Furthermore, local alignment of these intervals was performed against T2T-NIP reference (AGIS1.0) using Minimap2 (Li, 2021). Alignments were filtered to retain only those with sequence identity > 85% and aligned length > 10 Kb. Haplotype structures were manually curated and summarized across accessions. For barley and wheat, 76 and 18 genome assemblies were analyzed, respectively, alongside the reference genomes of 'B1K-04-12' (barley), and 'ZM366' (wheat) (Jayakodi et al., 2024; Jiao et al., 2025). Genomic intervals surrounding GBGC loci were extracted. In barley, alignment results were filtered using a threshold of length > 10 Kb and identity > 85% for barley and identity > 70% for wheat.

## Gramine-related gene expression

The barley (*Hordeum vulgare* cv. HOR9043) and *C. macrourus* was planted in a greenhouse with a 16 h photoperiod at 18 °C/15 °C (day/night). *Z. latifolia* and *O. sativa* were planted under natural filed condition. For exogenous jasmonic acid treatment, the 20-day plants were sprayed with 150 μM methyl jasmonic acid (MeJA, 392707, Sigma, Germany). The plant samples of leaves were collected with multiple courses (0, 0.5, 1, 2, 3, 4, 6, 9 and 12 hours). Total RNA under control or MeJA treatment were extracted by HiPure Plant RNA Plus Kit (R4150, Magen Biotechnology, Guangzhou, China). The cDNA was generated through using Hiscript II qRT Supermix for qPCR reagent (R223, Vazyme, Nanjing, China). The qRT-PCR assay was performed with SupRealQ Purple Universal SYBR qPCR Master Mix reagent (Q412, Vazyme, Nanjing, China) on CFX96 real-time system (Bio-Rad, USA), according to manufacturer's instructions. The reaction procedure was 95°C for 30s, 95°C for 10s and 60°C for 30s. The relative expression of genes was calculated by $2^{-\Delta\Delta Ct}$, using *EF1a, ACTB, OsActin* or *Zl18SRNA* as the internal reference. Primes used for constructs are listed in **Supplementary Table 7**.

## Gramine quantification

All samples were freeze-dried and ground into powder using a mixer mill. After the addition of 1000 $\mu$L of extraction solvent (precooled at -20 °C, methanol-water, 3:1), the samples were vortexed for 30 s, homogenized at 38 Hz for 4 min, and sonicated for 5 min in ice-water bath. The homogenate and sonicate circle were repeated for 3 times, followed by incubation at -20 °C for one hour and centrifugation at 12000 rpm and 4 °C for 15 min. An 80 $\mu$L aliquot of the clear supernatant was transferred to an auto-sampler vial for LC-MS/MS analysis. Another 50 $\mu$L of supernatant was diluted for 100 times for LC-MS/MS analysis. For gramine quantification, samples were analyzed via an ultrahigh performance liquid chromatography-tandem mass spectroscopy (UPLC-MS/MS)-based targeted method combined with a targeted metabolic profiling method. The UPLC analytical conditions were as follows: Waters ACQUITY UPLC BEH C18 (pore size 1.7µm, length 2.1 ×100mm); solvent system, water (0.1% formic acid) and methanol; gradient program, 0 min, 10% B; 2 min, 10% B; 5 min, 90% B; 7 min, 90% B;7.1 min, 10% B; 10 min, 10% B; flow rate, 300 $\mu$L/min; temperature, 35°C; injection volume: 1 µL. Targeted metabolic profiling analysis was performed using a scheduled multiple reaction monitoring (MRM) via Sciex QTRAP 6500+ triple quadrupole mass spectrometer, equipped with an electrospray ionization (ESI) interface. The ESI source operation parameters were as follows: curtain gas, 35 psi; ion spray voltage, 4500 V; source temperature, 550 °C; nebulizer gas, 55 psi; heater gas, 55 psi; declustering potential, 80 V; entrance potential, 10 V; collision cell exit potential, 13 V.

## Code availability

Source code for RIFinder are available at GitHub (https://github.com/Ne0tea/RIFinder) under Apache License 2.0.

## Data availability

Genome assemblies in this study are all downloaded from public data warehouses. The RI detection results for Poaceae species, including the initial output and filtered results from RIFinder, are publicly available on Zenodo at: 10.5281/zenodo.15804015.

## Acknowledgements

This work was supported by Young Scientists Fund of the National Natural Science Foundation of

China (No. 32300490) to D.W.

## Author contribution

D.W. conceived and initiated this study. Y.H. and S.Z collected the assemblies. S.Z and R.Z. performed quality control. Y.H, S.Z, Z.L, L.J. and K.Y. processed the RNA-seq data and analyzed the expression profile. Y.H., Y.L. and S.Z. performed the RIFinder evaluation based on simulated data. H.L. conduct ecological niche comparison. C.L., C.L. and Y.C. provided gramine quantification. C.H. and H.Z. provided samples and advice on experiments. Y.H. conducted the analysis of all major components. G.Z. and Q.Q. discussed the results. D.W. and L.F. supervised all analyses. Y.H. and D.W. wrote the manuscripts with the inputs from all co-authors. All authors discussed the results and commented on the manuscript.

## Competing interests

All authors declare no competing interests.

## Reference

Aguillon, S.M., Dodge, T.O., Preising, G.A., Schumer, M., 2022. Introgression. Current Biology 32, R865–R868.

Aharoni, A., Dixit, S., Jetter, R., Thoenes, E., van Arkel, G., Pereira, A., 2004. The SHINE Clade of AP2 Domain Transcription Factors Activates Wax Biosynthesis, Alters Cuticle Properties, and Confers Drought Tolerance when Overexpressed in Arabidopsis[W]. The Plant Cell 16, 2463–2480.

Anderson, E., Anderson, E., 1949. Introgressive hybridization. J. Wiley, New York.

Arnold, B.J., Huang, I.-T., Hanage, W.P., 2022. Horizontal gene transfer and adaptive evolution in bacteria. Nat Rev Microbiol 20, 206–218.

Bernard, A., Joubès, J., 2013. Arabidopsis cuticular waxes: Advances in synthesis, export and regulation. Progress in Lipid Research 52, 110–129.

Birchler, J.A., Yang, H., 2022. The multiple fates of gene duplications: Deletion, hypofunctionalization, subfunctionalization, neofunctionalization, dosage balance constraints, and neutral variation. The Plant Cell 34, 2466–2474.

Bjornson, S., Verbruggen, H., Upham, N.S., Steenwyk, J.L., 2024. Reticulate evolution: Detection and utility in the phylogenomics era. Molecular Phylogenetics and Evolution 201, 108197.

Brown, K.S., Trigo, J.R., 1995. Chapter 4 The Ecological Activity of Alkaloids. In: Cordell, G.A. (Ed.), The Alkaloids: Chemistry and Pharmacology. Academic Press, pp. 227–354.

Buchfink, B., Reuter, K., Drost, H.-G., 2021. Sensitive protein alignments at tree-of-life scale using DIAMOND. Nat Methods 18, 366–368.

Capella-Gutiérrez, S., Silla-Martínez, J.M., Gabaldón, T., 2009. trimAl: a tool for automated alignment trimming in large-scale phylogenetic analyses. Bioinformatics 25, 1972–1973.

Chen, Y., Guo, Y., Xie, X., Wang, Z., Miao, L., Yang, Z., Jiao, Y., Xie, C., Liu, J., Hu, Z., Xin, M., Yao, Y., Ni, Z., Sun, Q., Peng, H., Guo, W., 2023. Pangenome-based trajectories of intracellular gene transfers in Poaceae unveil high cumulation in Triticeae. Plant Physiology kiad319.

Corcuera, L.J., 1984. Effects of indole alkaloids from gramineae on aphids. Phytochemistry 23, 539–541.

Dias, S.L., Chuang, L., Liu, S., Seligmann, B., Brendel, F.L., Chavez, B.G., Hoffie, R.E., Hoffie, I., Kumlehn, J., Bültemeier, A., Wolf, J., Herde, M., Witte, C.-P., D'Auria, J.C., Franke, J., 2024. Biosynthesis of the allelopathic alkaloid gramine in barley by a cryptic oxidative rearrangement. Science 383, 1448–1454.

Dunning, L.T., Olofsson, J.K., Parisod, C., Choudhury, R.R., Moreno-Villena, J.J., Yang, Y., Dionora, J., Quick, W.P., Park, M., Bennetzen, J.L., Besnard, G., Nosil, P., Osborne, C.P., Christin, P.-A., 2019. Lateral transfers of large DNA fragments spread functional genes among grasses. Proceedings of the National Academy of Sciences 116, 4416–4425.

Edelman, N.B., Frandsen, P.B., Miyagi, M., Clavijo, B., Davey, J., Dikow, R.B., García-Accinelli, G., Van Belleghem, S.M., Patterson, N., Neafsey, D.E., Challis, R., Kumar, S., Moreira, G.R.P., Salazar, C., Chouteau, M., Counterman, B.A., Papa, R., Blaxter, M., Reed, R.D., Dasmahapatra, K.K., Kronforst, M., Joron, M., Jiggins, C.D., McMillan, W.O., Di Palma, F., Blumberg, A.J., Wakeley, J., Jaffe, D., Mallet, J., 2019a. Genomic architecture and introgression shape a butterfly radiation. Science 366, 594–599.

Edelman, N.B., Frandsen, P.B., Miyagi, M., Clavijo, B., Davey, J., Dikow, R.B., García-Accinelli, G., Van Belleghem, S.M., Patterson, N., Neafsey, D.E., Challis, R., Kumar, S., Moreira, G.R.P., Salazar, C., Chouteau, M., Counterman, B.A., Papa, R., Blaxter, M., Reed, R.D., Dasmahapatra, K.K., Kronforst, M., Joron, M., Jiggins, C.D., McMillan, W.O., Di Palma, F., Blumberg, A.J., Wakeley, J., Jaffe, D., Mallet, J., 2019b. Genomic architecture and introgression shape a butterfly radiation. Science 366, 594–599.

Edwards, S.V., Liu, L., Pearl, D.K., 2007. High-resolution species trees without concatenation. Proceedings of the National Academy of Sciences 104, 5936–5941.

Etten, J.V., Bhattacharya, D., 2020. Horizontal Gene Transfer in Eukaryotes: Not if, but How Much? Trends in Genetics 36, 915–925.

Fan, X., Qiu, H., Han, W., Wang, Y., Xu, D., Zhang, X., Bhattacharya, D., Ye, N., 2020. Phytoplankton pangenome reveals extensive prokaryotic horizontal gene transfer of diverse functions. Sci. Adv. 6, eaba0111.

Fontaine, M.C., Pease, J.B., Steele, A., Waterhouse, R.M., Neafsey, D.E., Sharakhov, I.V., Jiang, X., Hall, A.B., Catteruccia, F., Kakani, E., Mitchell, S.N., Wu, Y.-C., Smith, H.A., Love, R.R., Lawniczak, M.K., Slotman, M.A., Emrich, S.J., Hahn, M.W., Besansky, N.J., 2015. Extensive introgression in a malaria vector species complex revealed by phylogenomics. Science 347, 1258524.

Fu, X., Liu, C., Li, Y., Liao, S., Cheng, H., Tu, Y., Zhu, X., Chen, K., He, Y., Wang, G., 2021. The coordination of OsbZIP72 and OsMYBS2 with reverse roles regulates the transcription of *OsPsbS1* in rice. New Phytologist 229, 370–387.

Grass Phylogeny Working Group III, 2024. A nuclear phylogenomic tree of grasses (Poaceae) recovers current classification despite gene tree incongruence. New Phytologist nph.20263.

Guo, Z.-H., Ma, P.-F., Yang, G.-Q., Hu, J.-Y., Liu, Y.-L., Xia, E.-H., Zhong, M.-C., Zhao, L., Sun, G.-L., Xu, Y.-X., Zhao, Y.-J., Zhang, Y.-C., Zhang, Y.-X., Zhang, X.-M., Zhou, M.-Y., Guo, Y., Guo, C., Liu, J.-X., Ye, X.-Y., Chen, Y.-M., Yang, Y., Han, B., Lin, C.-S., Lu, Y., Li, D.-Z., 2019. Genome Sequences Provide Insights into the Reticulate Origin and Unique Traits of Woody Bamboos. Molecular Plant 12, 1353–1365.

He, F., Pasam, R., Shi, F., Kant, S., Keeble-Gagnere, G., Kay, P., Forrest, K., Fritz, A., Hucl, P., Wiebe, K., Knox, R., Cuthbert, R., Pozniak, C., Akhunova, A., Morrell, P.L., Davies, J.P., Webb, S.R., Spangenberg, G., Hayes, B., Daetwyler, H., Tibbits, J., Hayden, M., Akhunov, E., 2019. Exome sequencing highlights the role of wild-relative introgression in shaping the adaptive landscape of the wheat genome. Nat Genet 51, 896–904.

Hedrick, P.W., 2013. Adaptive introgression in animals: examples and comparison to new mutation and standing variation as sources of adaptive variation. Molecular Ecology 22, 4606–4618.

Huang, J., Yang, Lin, Yang, Liu, Wu, X., Cui, X., Zhang, L., Hui, J., Zhao, Y., Yang, H., Liu, S., Xu, Q., Pang, M., Guo, X., Cao, Y., Chen, Y., Ren, X., Lv, J., Yu, J., Ding, J., Xu, G., Wang, N., Wei, X., Lin, Q., Yuan, Y., Zhang, Xiaowei, Ma, C., Dai, C., Wang, P., Wang, Y., Cheng, F., Zeng, W., Palanivelu, R., Wu, H.-M., Zhang, Xiansheng, Cheung, A.Y., Duan, Q., 2023. Stigma receptors control intraspecies and interspecies barriers in Brassicaceae. Nature 614, 303–308.

Huang, W., Zhang, L., Columbus, J.T., Hu, Y., Zhao, Y., Tang, L., Guo, Z., Chen, W., McKain, M., Bartlett, M., Huang, C.-H., Li, D.-Z., Ge, S., Ma, H., 2022. A well-supported nuclear phylogeny of Poaceae and implications for the evolution of C4 photosynthesis. Molecular Plant 15, 755–777.

Huerta-Sánchez, E., Jin, X., Asan, Bianba, Z., Peter, B.M., Vinckenbosch, N., Liang, Y., Yi, X., He, M., Somel, M., Ni, P., Wang, B., Ou, X., Huasang, Luosang, J., Cuo, Z.X.P., Li, K., Gao, G., Yin, Y., Wang, W., Zhang, X., Xu, X., Yang, H., Li, Y., Wang, Jian, Wang, Jun, Nielsen, R., 2014. Altitude adaptation in Tibetans caused by introgression of Denisovan-like DNA. Nature 512, 194–197.

Husnik, F., McCutcheon, J.P., 2018. Functional horizontal gene transfer from bacteria to eukaryotes. Nat Rev Microbiol 16, 67–79.

Irwin, N.A.T., Pittis, A.A., Richards, T.A., Keeling, P.J., 2022. Systematic evaluation of horizontal gene transfer between eukaryotes and viruses. Nat Microbiol 7, 327–336.

Ishikawa, E., Kanai, S., Shinozawa, A., Hyakutake, M., Sue, M., 2024. Hordeum vulgare CYP76M57 catalyzes C shortening of tryptophan side chain by C–N bond rearrangement in gramine biosynthesis. The Plant Journal 118, 892–904.

Jensen, A., Horton, E.R., Amboko, J., Parke, S.-A., Hart, J.A., Tosi, A.J., Guschanski, K., Detwiler, K.M., 2024. Y chromosome introgression between deeply divergent primate species. Nat Commun 15, 10398.

Jensen, A., Swift, F., de Vries, D., Beck, R.M.D., Kuderna, L.F.K., Knauf, S., Chuma, I.S., Keyyu, J.D., Kitchener, A.C., Farh, K., Rogers, J., Marques-Bonet, T., Detwiler, K.M., Roos, C., Guschanski, K., 2023. Complex Evolutionary History With Extensive Ancestral Gene Flow in an African Primate Radiation. Molecular Biology and Evolution 40, msad247.

Jing, C.-Y., Zhang, F.-M., Wang, X.-H., Wang, M.-X., Zhou, L., Cai, Z., Han, J.-D., Geng, M.-F., Yu, W.-H., Jiao, Z.-H., Huang, L., Liu, R., Zheng, X.-M., Meng, Q.-L., Ren, N.-N., Zhang, H.-X., Du, Y.-S., Wang, X., Qiang, C.-G., Zou, X.-H., Gaut, B.S., Ge, S., 2023. Multiple domestications of Asian rice. Nat. Plants 9, 1221–1235.

Jonczyk, R., Schmidt, H., Osterrieder, A., Fiesselmann, A., Schullehner, K., Haslbeck, M., Sicker, D., Hofmann, D., Yalpani, N., Simmons, C., Frey, M., Gierl, A., 2008. Elucidation of the Final Reactions of DIMBOA-Glucoside Biosynthesis in Maize: Characterization of Bx6 and Bx7. Plant Physiology 146, 1053–1063.

Kalyaanamoorthy, S., Minh, B.Q., Wong, T.K.F., von Haeseler, A., Jermiin, L.S., 2017. ModelFinder: fast model selection for accurate phylogenetic estimates. Nat Methods 14, 587–589.

Keeling, P.J., 2024. Horizontal gene transfer in eukaryotes: aligning theory with data. Nat Rev Genet 25, 416–430.

Keeling, P.J., Palmer, J.D., 2008. Horizontal gene transfer in eukaryotic evolution. Nat Rev Genet 9, 605–618.

Kim, D., Pertea, G., Trapnell, C., Pimentel, H., Kelley, R., Salzberg, S.L., 2013. TopHat2: accurate alignment of transcriptomes in the presence of insertions, deletions and gene fusions. Genome Biology 14, R36.

Kishino, H., Miyata, T., Hasegawa, M., 1990. Maximum likelihood inference of protein phylogeny and the origin of chloroplasts. J Mol Evol 31, 151–160.

Kloub, L., Gosselin, S., Fullmer, M., Graf, J., Gogarten, J.P., Bansal, M.S., 2021. Systematic Detection of Large-Scale Multigene Horizontal Transfer in Prokaryotes. Molecular Biology and Evolution 38, 2639–2659.

Kominek, J., Doering, D.T., Opulente, D.A., Shen, X.-X., Zhou, X., DeVirgilio, J., Hulfachor, A.B., Groenewald, M., Mcgee, M.A., Karlen, S.D., Kurtzman, C.P., Rokas, A., Hittinger, C.T., 2019. Eukaryotic Acquisition of a Bacterial Operon. Cell 176, 1356-1366.e10.

Kulmuni, J., Butlin, R.K., Lucek, K., Savolainen, V., Westram, A.M., 2020. Towards the completion of speciation: the evolution of reproductive isolation beyond the first barriers. Philosophical Transactions of the Royal Society B.

Lan, Z., Song, Z., Wang, Z., Li, L., Liu, Y., Zhi, S., Wang, R., Wang, J., Li, Q., Bleckmann, A., Zhang, L., Dresselhaus, T., Dong, J., Gu, H., Zhong, S., Qu, L.-J., 2023. Antagonistic RALF peptides control an intergeneric hybridization barrier on Brassicaceae stigmas. Cell 186, 4773-4787.e12.

Langfelder, P., Horvath, S., 2008. WGCNA: an R package for weighted correlation network analysis. BMC Bioinformatics 9, 559.

Leete, E., Minich, M.L., 1977. Biosynthesis of gramine in Phalaris arundinacea. Phytochemistry 16, 149–150.

Letunic, I., Bork, P., 2021. Interactive Tree Of Life (iTOL) v5: an online tool for phylogenetic tree display and annotation. Nucleic Acids Res 49, W293–W296.

Li, Y., Li, X., Nie, S., Zhang, M., Yang, Q., Xu, W., Duan, Y., Wang, X., 2024. Reticulate evolution of the tertiary relict Osmanthus. The Plant Journal 117, 145–160.

Li, Y., Liu, Z., Liu, C., Shi, Z., Pang, L., Chen, C., Chen, Y., Pan, R., Zhou, W., Chen, X., Rokas, A., Huang, J., Shen, X.-X., 2022a. HGT is widespread in insects and contributes to male courtship in lepidopterans. Cell 0.

Li, Y., Liu, Z., Liu, C., Shi, Z., Pang, L., Chen, C., Chen, Y., Pan, R., Zhou, W., Chen, X., Rokas, A., Huang, J., Shen, X.-X., 2022b. HGT is widespread in insects and contributes to male courtship in lepidopterans. Cell 185, 2975-2987.e10.

Liao, Y., Smyth, G.K., Shi, W., 2014. featureCounts: an efficient general purpose program for assigning sequence reads to genomic features. Bioinformatics 30, 923–930.

Lin, T., Zhu, G., Zhang, J., Xu, X., Yu, Q., Zheng, Z., Zhang, Z., Lun, Y., Li, S., Wang, X., Huang, Z., Li, Junming, Zhang, C., Wang, T., Zhang, Yuyang, Wang, A., Zhang, Yancong, Lin, K., Li, C., Xiong, G., Xue, Y., Mazzucato, A., Causse, M., Fei, Z., Giovannoni, J.J., Chetelat, R.T., Zamir, D., Städler, T., Li, Jingfu, Ye, Z., Du, Y., Huang, S., 2014. Genomic analyses provide insights into the history of tomato breeding. Nat Genet 46, 1220–1226.

Lu, A., Wang, T., Hui, H., Wei, X., Cui, W., Zhou, C., Li, H., Wang, Z., Guo, J., Ma, D., Wang, Q., 2019. Natural Products for Drug Discovery: Discovery of Gramines as Novel Agents against a Plant Virus. J. Agric. Food Chem. 67, 2148–2156.

Ma, J., Wang, S., Zhu, X., Sun, G., Chang, G., Li, L., Hu, X., Zhang, S., Zhou, Y., Song, C.-P., Huang, J., 2022. Major episodes of horizontal gene transfer drove the evolution of land plants. Molecular Plant 15, 857–871.

Ma, P.-F., Liu, Y.-L., Guo, C., Jin, G., Guo, Z.-H., Mao, L., Yang, Y.-Z., Niu, L.-Z., Wang, Y.-J., Clark, L.G., Kellogg, E.A., Xu, Z.-C., Ye, X.-Y., Liu, J.-X., Zhou, M.-Y., Luo, Y., Yang, Y., Soltis, D.E., Bennetzen, J.L., Soltis, P.S., Li, D.-Z., 2024. Genome assemblies of 11 bamboo species highlight diversification induced by dynamic subgenome dominance. Nat Genet 56, 710–720.

Magallón, S., Gómez-Acevedo, S., Sánchez-Reyes, L.L., Hernández-Hernández, T., 2015. A metacalibrated time-tree documents the early rise of flowering plant phylogenetic diversity. New Phytologist 207, 437–453.

Mahelka, V., Krak, K., Fehrer, J., Caklová, P., Nagy Nejedlá, M., Čegan, R., Kopecký, D., Šafář, J., 2021. A Panicum-derived chromosomal segment captured by Hordeum a few million years ago preserves a set of stress-related genes. The Plant Journal 105, 1141–1164.

Mahelka, V., Krak, K., Kopecký, D., Fehrer, J., Šafář, J., Bartoš, J., Hobza, R., Blavet, N., Blattner, F.R., 2017. Multiple horizontal transfers of nuclear ribosomal genes between phylogenetically distinct grass lineages. Proc. Natl. Acad. Sci. U.S.A. 114, 1726–1731.

Marques, D.A., Meier, J.I., Seehausen, O., 2019. A Combinatorial View on Speciation and Adaptive Radiation. Trends in Ecology & Evolution 34, 531–544.

Meng, X., Li, L., Pascual, J., Rahikainen, M., Yi, C., Jost, R., He, C., Fournier-Level, A., Borevitz, J., Kangasjärvi, S., Whelan, J., Berkowitz, O., 2022. GWAS on multiple traits identifies mitochondrial ACONITASE3 as important for acclimation to submergence stress. Plant Physiol 188, 2039–2058.

Minh, B.Q., Schmidt, H.A., Chernomor, O., Schrempf, D., Woodhams, M.D., von Haeseler, A., Lanfear, R., 2020. IQ-TREE 2: New models and efficient methods for phylogenetic inference in the genomic era. Molecular Biology and Evolution 37, 1530–1534.

Moran, B.M., Payne, C., Langdon, Q., Powell, D.L., Brandvain, Y., Schumer, M., 2021. The genomic consequences of hybridization. eLife 10, e69016.

Muzellec, B., Teleńczuk, M., Cabeli, V., Andreux, M., 2023. PyDESeq2: a python package for bulk RNA-seq differential expression analysis. Bioinformatics 39, btad547.

Nosil, P., Harmon, L.J., Seehausen, O., 2009. Ecological explanations for (incomplete) speciation. Trends in Ecology & Evolution 24, 145–156.

Oziolor, E.M., Reid, N.M., Yair, S., Lee, K.M., Guberman VerPloeg, S., Bruns, P.C., Shaw, J.R.,

Whitehead, A., Matson, C.W., 2019. Adaptive introgression enables evolutionary rescue from extreme environmental pollution. Science 364, 455–457.

Pascual, J., Rahikainen, M., Angeleri, M., Alegre, S., Gossens, R., Shapiguzov, A., Heinonen, A., Trotta, A., Durian, G., Winter, Z., Sinkkonen, J., Kangasjärvi, J., Whelan, J., Kangasjärvi, S., 2021. ACONITASE 3 is part of theANAC017 transcription factor-dependent mitochondrial dysfunction response. Plant Physiol 186, 1859–1877.

Peccoud, J., Loiseau, V., Cordaux, R., Gilbert, C., 2017. Massive horizontal transfer of transposable elements in insects. Proceedings of the National Academy of Sciences 114, 4721–4726.

Pontarp, M., Lundberg, P., Ripa, J., 2024. The succession of ecological divergence and reproductive isolation in adaptive radiations. Journal of Theoretical Biology 587, 111819.

Ramírez-González, R.H., Borrill, P., Lang, D., Harrington, S.A., Brinton, J., Venturini, L., Davey, M., Jacobs, J., van Ex, F., Pasha, A., Khedikar, Y., Robinson, S.J., Cory, A.T., Florio, T., Concia, L., Juery, C., Schoonbeek, H., Steuernagel, B., Xiang, D., Ridout, C.J., Chalhoub, B., Mayer, K.F.X., Benhamed, M., Latrasse, D., Bendahmane, A., International Wheat Genome Sequencing Consortium, Wulff, B.B.H., Appels, R., Tiwari, V., Datla, R., Choulet, F., Pozniak, C.J., Provart, N.J., Sharpe, A.G., Paux, E., Spannagl, M., Bräutigam, A., Uauy, C., 2018. The transcriptional landscape of polyploid wheat. Science 361, eaar6089.

Rozewicki, J., Li, S., Amada, K.M., Standley, D.M., Katoh, K., 2019. MAFFT-DASH: integrated protein sequence and structural alignment. Nucleic Acids Research 47, W5–W10.

Seehausen, O., Takimoto, G., Roy, D., Jokela, J., 2008. Speciation reversal and biodiversity dynamics with hybridization in changing environments. Molecular Ecology 17, 30–44.

Selinski, J., Hartmann, A., Deckers-Hebestreit, G., Day, D.A., Whelan, J., Scheibe, R., 2018. Alternative Oxidase Isoforms Are Differentially Activated by Tricarboxylic Acid Cycle Intermediates. Plant Physiol 176, 1423–1432.

Shannon, P., Markiel, A., Ozier, O., Baliga, N.S., Wang, J.T., Ramage, D., Amin, N., Schwikowski, B., Ideker, T., 2003. Cytoscape: a software environment for integrated models of biomolecular interaction networks. Genome Res 13, 2498–2504.

Shao, Y., Zhou, L., Li, F., Zhao, L., Zhang, B.-L., Shao, F., Chen, J.-W., Chen, C.-Y., Bi, X., Zhuang, X.-L., Zhu, H.-L., Hu, J., Sun, Z., Li, X., Wang, D., Rivas-González, I., Wang, S., Wang, Y.-M., Chen, W., Li, G., Lu, H.-M., Liu, Y., Kuderna, L.F.K., Farh, K.K.-H., Fan, P.-F., Yu, L., Li, M., Liu, Z.-J., Tiley, G.P., Yoder, A.D., Roos, C., Hayakawa, T., Marques-Bonet, T., Rogers, J., Stenson, P.D., Cooper, D.N., Schierup, M.H., Yao, Y.-G., Zhang, Y.-P., Wang, W., Qi, X.-G., Zhang, G., Wu, D.-D., 2023. Phylogenomic analyses provide insights into primate evolution. Science 380, 913–924.

Soreng, R.J., Peterson, P.M., Romaschenko, K., Davidse, G., Teisher, J.K., Clark, L.G., Barberá, P., Gillespie, L.J., Zuloaga, F.O., 2017. A worldwide phylogenetic classification of the Poaceae (Gramineae) II: An update and a comparison of two 2015 classifications. Journal of Systematics and Evolution 55, 259–290.

Soucy, S.M., Huang, J., Gogarten, J.P., 2015. Horizontal gene transfer: building the web of life. Nat Rev Genet 16, 472–482.

Steenwyk, J.L., Li, Y., Zhou, X., Shen, X.-X., Rokas, A., 2023. Incongruence in the phylogenomics era. Nat Rev Genet 24, 834–850.

Stull, G.W., Pham, K.K., Soltis, P.S., Soltis, D.E., 2023. Deep reticulation: the long legacy of hybridization in vascular plant evolution. The Plant Journal 114, 743–766.

Sun, X.-Q., Zhang, M.-X., Yu, J.-Y., Jin, Y., Ling, B., Du, J.-P., Li, G.-H., Qin, Q.-M., Cai, Q.-N., 2013. Glutathione S-Transferase of Brown Planthoppers (Nilaparvata lugens) Is Essential for Their Adaptation to Gramine-Containing Host Plants. PLOS ONE 8, e64026.

Supek, F., Bošnjak, M., Škunca, N., Šmuc, T., 2011. REVIGO Summarizes and Visualizes Long Lists of Gene Ontology Terms. PLOS ONE 6, e21800.

Tang, H., Krishnakumar, V., Zeng, X., Xu, Z., Taranto, A., Lomas, J.S., Zhang, Y., Huang, Y., Wang, Y., Yim, W.C., Zhang, J., Zhang, X., 2024. JCVI: A versatile toolkit for comparative genomics analysis. iMeta 3, e211.

Than, C., Ruths, D., Nakhleh, L., 2008. PhyloNet: a software package for analyzing and reconstructing reticulate evolutionary relationships. BMC Bioinformatics 9, 322.

Wang, H., Sun, S., Ge, W., Zhao, L., Hou, B., Wang, K., Lyu, Z., Chen, L., Xu, S., Guo, J., Li, M., Su, P., Li, Xuefeng, Wang, G., Bo, C., Fang, X., Zhuang, W., Cheng, X., Wu, J., Dong, L., Chen, W., Li, W., Xiao, G., Zhao, J., Hao, Y., Xu, Y., Gao, Y., Liu, W., Liu, Y., Yin, H., Li, J., Li, Xiang, Zhao, Y., Wang, X., Ni, F., Ma, X., Li, A., Xu, S.S., Bai, G., Nevo, E., Gao, C., Ohm, H., Kong, L., 2020. Horizontal gene transfer of Fhb7 from fungus underlies Fusarium head blight resistance in wheat. Science 368, eaba5435.

Wei, X., Chen, M., Zhang, Q., Gong, J., Liu, J., Yong, K., Wang, Q., Fan, J., Chen, S., Hua, H., Luo, Z., Zhao, X., Wang, X., Li, W., Cong, J., Yu, X., Wang, Z., Huang, R., Chen, J., Zhou, X., Qiu, J., Xu, P., Murray, J., Wang, H., Xu, Y., Xu, C., Xu, G., Yang, J., Han, B., Huang, X., 2024. Genomic investigation of 18,421 lines reveals the genetic architecture of rice. Science 385, eadm8762.

Wei, X., Qiu, J., Yong, K., Fan, J., Zhang, Q., Hua, H., Liu, J., Wang, Q., Olsen, K.M., Han, B., Huang, X., 2021. A quantitative genomics map of rice provides genetic insights and guides breeding. Nat Genet 53, 243–253.

Whitney, K.D., Randell, R.A., Rieseberg, L.H., 2010. Adaptive introgression of abiotic tolerance traits in the sunflower Helianthus annuus. New Phytologist 187, 230–239.

Wu, D., Hu, Y., Akashi, S., Nojiri, H., Guo, L., Ye, C.-Y., Zhu, Q.-H., Okada, K., Fan, L., 2022a. Lateral transfers lead to the birth of momilactone biosynthetic gene clusters in grass. The Plant Journal 111, 1354–1367.

Wu, D., Jiang, B., Ye, C.-Y., Timko, M.P., Fan, L., 2022b. Horizontal transfer and evolution of the biosynthetic gene cluster for benzoxazinoids in plants. Plant Communications, Special Issue on Rice Functional Genomics 3, 100320.

Wu, D., Xie, L., Sun, Y., Huang, Y., Jia, L., Dong, C., Shen, E., Ye, C.-Y., Qian, Q., Fan, L., 2023. A syntelog-based pan-genome provides insights into rice domestication and de-domestication. Genome Biology 24, 179.

Wu, J.-J., Deng, Q.-W., Qiu, Y.-Y., Liu, C., Lin, C.-F., Ru, Y.-L., Sun, Y., Lai, J., Liu, L.-X., Shen, X.-X., Pan, R., Zhao, Y.-P., 2024. Post-transfer adaptation of HGT-acquired genes and contribution to guanine metabolic diversification in land plants. New Phytologist 244, 694–707.

Wu, Y., You, H.-L., Li, X.-Q., 2018. Dinosaur-associated Poaceae epidermis and phytoliths from the Early Cretaceous of China. National Science Review 5, 721–727.

Xia, J., Guo, Z., Yang, Z., Han, H., Wang, S., Xu, H., Yang, X., Yang, F., Wu, Q., Xie, W., Zhou, X., Dermauw, W., Turlings, T.C.J., Zhang, Y., 2021. Whitefly hijacks a plant detoxification gene that neutralizes plant toxins. Cell 184, 1693-1705.e17.

Yang, X., Yu, H., Sun, W., Ding, L., Li, J., Cheema, J., Ramirez-Gonzalez, R., Zhao, X., Martín, A.C., Lu, F., Liu, B., Uauy, C., Ding, Y., Zhang, H., 2021. Wheat in vivo RNA structure landscape reveals a prevalent role of RNA structure in modulating translational subgenome expression asymmetry. Genome Biology 22, 326.

Yang, Z., 2007. PAML 4: Phylogenetic Analysis by Maximum Likelihood. Molecular Biology and Evolution 24, 1586–1591.

Yang, Z., Wafula, E.K., Kim, G., Shahid, S., McNeal, J.R., Ralph, P.E., Timilsena, P.R., Yu, W., Kelly, E.A., Zhang, H., Person, T.N., Altman, N.S., Axtell, M.J., Westwood, J.H., dePamphilis, C.W., 2019. Convergent horizontal gene transfer and cross-talk of mobile nucleic acids in parasitic plants. Nat. Plants 5, 991–1001.

Yeasmin, L., Ali, Md.N., Gantait, S., Chakraborty, S., 2015. Bamboo: an overview on its genetic diversity and characterization. 3 Biotech 5, 1–11.

Yoon, J., Cho, L.-H., Yang, W., Pasriga, R., Wu, Y., Hong, W.-J., Bureau, C., Wi, S.J., Zhang, T., Wang, R., Zhang, D., Jung, K.-H., Park, K.Y., Périn, C., Zhao, Y., An, G., 2020. Homeobox transcription factor OsZHD2 promotes root meristem activity in rice by inducing ethylene biosynthesis. Journal of Experimental Botany 71, 5348–5364.

Yoshida, S., Maruyama, S., Nozaki, H., Shirasu, K., 2010. Horizontal Gene Transfer by the Parasitic Plant Striga hermonthica. Science 328, 1128–1128.

Zhang, C., Nielsen, R., Mirarab, S., 2025. CASTER: Direct species tree inference from whole-genome alignments. Science 0, eadk9688.

Zhang, L., Zhu, X., Zhao, Y., Guo, J., Zhang, T., Huang, W., Huang, J., Hu, Y., Huang, C.-H., Ma, H., 2022. Phylotranscriptomics Resolves the Phylogeny of Pooideae and Uncovers Factors for Their Adaptive Evolution. Mol Biol Evol 39.

Zhao, J.-X., Wang, S., Liu, Jiazhi, Jiang, X.-D., Wen, J., Suo, Z.-Q., Liu, Jie, Zhong, M.-C., Wang, Q., Gu, Z., Liu, C., Deng, Y., Hu, J.-Y., Li, D.-Z., 2023. A comparative full-length transcriptomic resource provides insight into the perennial monocarpic mass flowering. The Plant Journal 116, 1842–1855.

Aiello-Lammens, M.E., Boria, R.A., Radosavljevic, A., Vilela, B., Anderson, R.P., 2015. spThin: an R package for spatial thinning of species occurrence records for use in ecological niche models. Ecography 38, 541–545.

Di Cola, V., Broennimann, O., Petitpierre, B., Breiner, F.T., D'Amen, M., Randin, C., Engler, R., Pottier, J., Pio, D., Dubuis, A., Pellissier, L., Mateo, R.G., Hordijk, W., Salamin, N., Guisan, A., 2017. ecospat: an R package to support spatial analyses and modeling of species niches and distributions. Ecography 40, 774–787.

Fick, S.E., Hijmans, R.J., 2017. WorldClim 2: new 1-km spatial resolution climate surfaces for global land areas. International Journal of Climatology 37, 4302–4315.

Schoener, T.W., 1968. The Anolis Lizards of Bimini: Resource Partitioning in a Complex Fauna. Ecology 49, 704–726.

Warren, D.L., Glor, R.E., Turelli, M., 2008. Environmental Niche Equivalency Versus Conservatism: Quantitative Approaches to Niche Evolution. Evolution 62, 2868–2883.

Das, N., Bhattacharya, S., Bhattacharyya, S., Maiti, M.K., 2018. Expression of rice MATE family

transporter OsMATE2 modulates arsenic accumulation in tobacco and rice. Plant Mol Biol 98, 101–120.

Nagar, P., Sharma, N., Jain, M., Sharma, G., Prasad, M., Mustafiz, A., 2022. OsPSKR15, a phytosulfokine receptor from rice enhances abscisic acid response and drought stress tolerance. Physiologia Plantarum 174, e13569.

Zdobnov, E.M., Apweiler, R., 2001. InterProScan – an integration platform for the signature-recognition methods in InterPro. Bioinformatics 17, 847–848.

*Adscita splendens* subsp. *splendens* in GBIF Secretariat (2023). GBIF Backbone Taxonomy. Checklist dataset https://doi.org/10.15468/39omei accessed via GBIF.org on 2025-07-02.

*Cleistogenes songorica* (Roshev.) Ohwi in GBIF Secretariat (2023). GBIF Backbone Taxonomy. Checklist dataset https://doi.org/10.15468/39omei accessed via GBIF.org on 2025-07-02.

*Zoysia japonica* Steud. in GBIF Secretariat (2023). GBIF Backbone Taxonomy. Checklist dataset https://doi.org/10.15468/39omei accessed via GBIF.org on 2025-07-02.

Hahn, M.W., Hibbins, M.S., 2019. A Three-Sample Test for Introgression. Molecular Biology and Evolution 36, 2878–2882.

Suvorov, A., Kim, B.Y., Wang, J., Armstrong, E.E., Peede, D., D'Agostino, E.R.R., Price, D.K., Waddell, P.J., Lang, M., Courtier-Orgogozo, V., David, J.R., Petrov, D., Matute, D.R., Schrider, D.R., Comeault, A.A., 2022. Widespread introgression across a phylogeny of 155 Drosophila genomes. Current Biology 32, 111-123.e5.

Sjöstrand, J., Tofigh, A., Daubin, V., Arvestad, L., Sennblad, B., Lagergren, J., 2014. A Bayesian Method for Analyzing Lateral Gene Transfer. Systematic Biology 63, 409–420.

Yu, Y., Zhang, H., Long, Y., Shu, Y., Zhai, J., 2022. Plant Public RNA-seq Database: a comprehensive online database for expression analysis of 45 000 plant public RNA-Seq libraries. Plant Biotechnology Journal 20, 806–808.

Li, H., Feng, X., Chu, C., 2020. The design and construction of reference pangenome graphs with minigraph. Genome Biology 21, 265.

Wick, R.R., Schultz, M.B., Zobel, J., Holt, K.E., 2015. Bandage: interactive visualization of de novo genome assemblies. Bioinformatics 31, 3350–3352.

Xie, L., Huang, Y., Huang, W., Shang, L., Sun, Y., Chen, Q., Bi, S., Suo, M., Zhang, S., Yang, C., Zheng, X., Jin, W., Qian, Q., Fan, L., Wu, D., 2024. Genetic diversity and evolution of rice centromeres.

Li, H., 2021. New strategies to improve minimap2 alignment accuracy. Bioinformatics 37, 4572–4574.

Shang, L., He, W., Wang, T., Yang, Y., Xu, Q., Zhao, X., Yang, L., Zhang, H., Li, X., Lv, Y., Chen, W., Cao, S., Wang, X., Zhang, Bin, Liu, X., Yu, X., He, H., Wei, H., Leng, Y., Shi, C., Guo, M., Zhang, Z., Zhang, Bintao, Yuan, Q., Qian, H., Cao, X., Cui, Y., Zhang, Q., Dai, X., Liu, C., Guo, L., Zhou, Y., Zheng, X., Ruan, J., Cheng, Z., Pan, W., Qian, Q., 2023. A complete assembly of the rice Nipponbare reference genome. Molecular Plant 16, 1232–1236.

Jayakodi, M., Lu, Q., Pidon, H., Rabanus-Wallace, M.T., Bayer, M., Lux, T., Guo, Y., Jaegle, B., Badea, A., Bekele, W., Brar, G.S., Braune, K., Bunk, B., Chalmers, K.J., Chapman, B., Jørgensen, M.E., Feng, J.-W., Feser, M., Fiebig, A., Gundlach, H., Guo, W., Haberer, G., Hansson, M., Himmelbach, A., Hoffie, I., Hoffie, R.E., Hu, H., Isobe, S., König, P., Kale, S.M., Kamal, N.,

Keeble-Gagnère, G., Keller, B., Knauft, M., Koppolu, R., Krattinger, S.G., Kumlehn, J., Langridge, P., Li, C., Marone, M.P., Maurer, A., Mayer, K.F.X., Melzer, M., Muehlbauer, G.J., Murozuka, E., Padmarasu, S., Perovic, D., Pillen, K., Pin, P.A., Pozniak, C.J., Ramsay, L., Pedas, P.R., Rutten, T., Sakuma, S., Sato, K., Schüler, D., Schmutzer, T., Scholz, U., Schreiber, M., Shirasawa, K., Simpson, C., Skadhauge, B., Spannagl, M., Steffenson, B.J., Thomsen, H.C., Tibbits, J.F., Nielsen, M.T.S., Trautewig, C., Vequaud, D., Voss, C., Wang, P., Waugh, R., Westcott, S., Rasmussen, M.W., Zhang, R., Zhang, X.-Q., Wicker, T., Dockter, C., Mascher, M., Stein, N., 2024. Structural variation in the pangenome of wild and domesticated barley. Nature 636, 654–662.

Jiao, C., Xie, X., Hao, C., Chen, L., Xie, Y., Garg, V., Zhao, L., Wang, Z., Zhang, Y., Li, T., Fu, J., Chitikineni, A., Hou, J., Liu, H., Dwivedi, G., Liu, X., Jia, J., Mao, L., Wang, X., Appels, R., Varshney, R.K., Guo, W., Zhang, X., 2025. Pan-genome bridges wheat structural variations with habitat and breeding. Nature 637, 384–393.

# Figure Legends

**Figure 1 | RIFinder workflow and performance evaluation.**

**a**, Overview of RIFinder processing steps. Tips from closely related species are marked with circles of the same color. Phylogeny trees of different taxes are shown in a gray background. Red triangles mark candidate RI nodes detected via topological conflict analysis. **b**, Performance evaluation on simulated datasets. Upper panels represent distinct phylogenetic scenarios, and bar charts display precision (left, solid lines) and recall (right, right dashed lines) metrics. The x-axis indicates the gene transfer probability parameter in simulations.

**Figure 2 | Widespread remote introgression events between PACMAD and BOP lineages in Poaceae.**

**a**, RI events as donor (left half pie) and acceptor (right half pie) across the grass family. Pie diameter at the internal node shows the transfer frequency and the color denotes the donor/acceptor taxonomy. Bar height indicates the frequency of RI events in leaves. Major crops are annotated with red dots at leaf nodes, while orphan crops with light red. Poaceae subfamilies are labeled with silhouettes of representative species (images obtained from BioRender, https://biorender.com). Nodes with bootstrap support < 95 are marked with red stars. BAM, Bambusoideae; ORY, Oryzoideae, POO, Pooideae; CHL, Chloridoideae; PAN, Panicoideae; Arun, Arundinarieae; Brach, Brachypodieae; Andro, Andropogoneae; Cyn, Cynodonteae; Erag, Eragrostideae. **b**, Relationship between RI events as donors and acceptors. Node colors correspond to species taxonomy. **c**, Distribution of RI time estimates in rice and wheat, with the times for *Bx6* and *Bx6-like* RI events highlighted in red. **d**, A maximum-likelihood phylogenetic tree of *Bx6* homologs using protein sequences from grasses, with *P. latifolius* serving as an outgroup. Nodes with bootstrap values < 60 are colored in black, while those with values 60-80 are colored in light blue. Leaf background colors correspond to species taxonomic classification. The *Bx6-like* branches of RI origin are highlighted in red.

**Figure 3 | Functional characterization of RI-originated genes in the grass family.**

**a-d**, Top 10 enriched gene families of RI acceptor genes in rice (*O. sativa*), wheat (*T. aestivum*), maize (*Zea mays*), and tef (*Eragrostis tef*). Color gradient indicates the statistical significance (Fisher's exact test, *p*-adjusted) of gene family enrichment ratios relative to the whole-genome background. **e**, Expression of 54 RI derived genes in rice across tissues and stress treatments. Chromosomal ideograms above the heatmap annotate the genomic loci of these RI genes. Colored rectangles denote gene categories: gray for native orthologs (connected to RI genes via dashed lines), blue for cloned abiotic stress-resistance genes, and red for biotic stress-resistance genes.

Stars mark genes within selective sweep regions (Jing et al., 2023). **f**, Expression of 35 RI-originated genes and their corresponding native copies in wheat across tissues and experimental treatments. Subgenome-specific absences are marked by hollow rectangles with dashed borders. JA, Jasmonates; Minc, Rice blast; RSV, Rice stripe virus; BPH, Brown planthopper; Xt, *Xanthomonas translucens*; Zt, *Zymoseptoria tritici*; Stripe, Stripe rust; Powdery, Powdery Mildew; Cp, *Claviceps purpurea*; Fg, *Fusarium graminearum*; Have, *Heterodera avenae*.

**Figure 4 | Evolution and functional implication of a Triticeae-derived 30-Kbp segment in *Cleistogenes songorica*.**

**a**, Representative plants of *C. songorica* (Cson, left) and *A. splendens* (Aspl, right). Credits: Kherlenchimeg Nyamsuren (Cson) and Soreng, R.J. (Aspl). **b**, Maximum-likelihood phylogenetic trees of *RIGs*. *P. latifolius* and *A. comosus* serve as outgroups. Branches are colored by species taxonomic classification. Nodes with bootstrap values < 60 are shown in black and values 60-80 in light blue. Nested topological structures are magnified in gray blocks, with *RIGs* highlighted. Rectangular blocks at terminal nodes denote syntenic group (SG) assignment: same-colored blocks indicate genes within the same SG, and gray blocks represent singleton SGs. **c**, Genomic synteny of the RI-originated segment across grass families. *RIG* homologs are colored and connected by red lines; other genes are gray. Chromosome labels are colored by species taxonomy. Orientations of *RIGs* and their native copies are shown with arrows. **d**, Nucleotide divergence between homologous segments in Aspl and Cson, calculated in 200-bp sliding windows. Statistical significance of distribution differences was tested via two-tailed *t*-test. **e**, Expression profiles of *RIGs* and *HRIGs* under high-temperature and drought stress. **f**, Geographical distribution of *Z. japonica*, Cson, and Aspl. Data source: Global Biodiversity Information Facility (GBIF; https://doi.org/10.15468/dl.uveqts). **g**, Ecological niche differentiation across two climatic gradients for *Z. japonica*, Cson, and Aspl.

**Figure 5 | RI-mediated evolution of gramine biosynthesis genes.**

**a**, Genomic locus of gramine biosynthesis gene cluster (GBGC) in wild barley, *H. vulgare* subsp. *spontaneum* var. B1K-04-12, and biosynthesis pathway of gramine from tryptophan. **b**, A maximum-likelihood phylogenetic tree of *AMIS* homologs in grasses. Core P450 genes of Clade 2 (from *CYP76M5* to *CYP76M8*) from the ORY subfamily are collapsed, with leaf node numbers labeled. Background colors denote grass subfamilies. RI-related clade is highlighted in red. Rectangular blocks at terminal nodes indicate SGs. **c**, Evolutionary distribution of GBGC genes in grass genomes. The numbers between *AMIS* and *NMT* represent their genomic distances. Genes flanking the slash are located on two different chromosomes. **d** and **f**, Location and gene element around GBGCs in *O. sativa* var. NIP and *T. aestivum* var. Chinese spring. **e**, Graph of rice pan-genome at the GBGC locus. Two major haplotypes are represented in green and blue. Circular plot below illustrates the proportional distribution of these haplotypes across different rice groups. African, *Oryza barthii* and *Oryza glaberrima*; Or-12, group 1/2 of *O. rufipogen*; XI, *O. sativa* ssp. *xian/indica*; Or-3, group 3 of *O. rufipogen*; GJ, *O. sativa* ssp. *geng/japonica*. **g**, *TaAMISa*, *TaELL1* and *TaAMISb* expression profiles in the spike of wheat under *Fusarium* head blight treatment. **h**, Extracted ion chromatograms (EICs) showing gramine (m/z = 175.12, product) detection in *C. macrourus* and *H. vulgare* (cv. 7552), based on HPLC-MS analysis.

# Figure 1

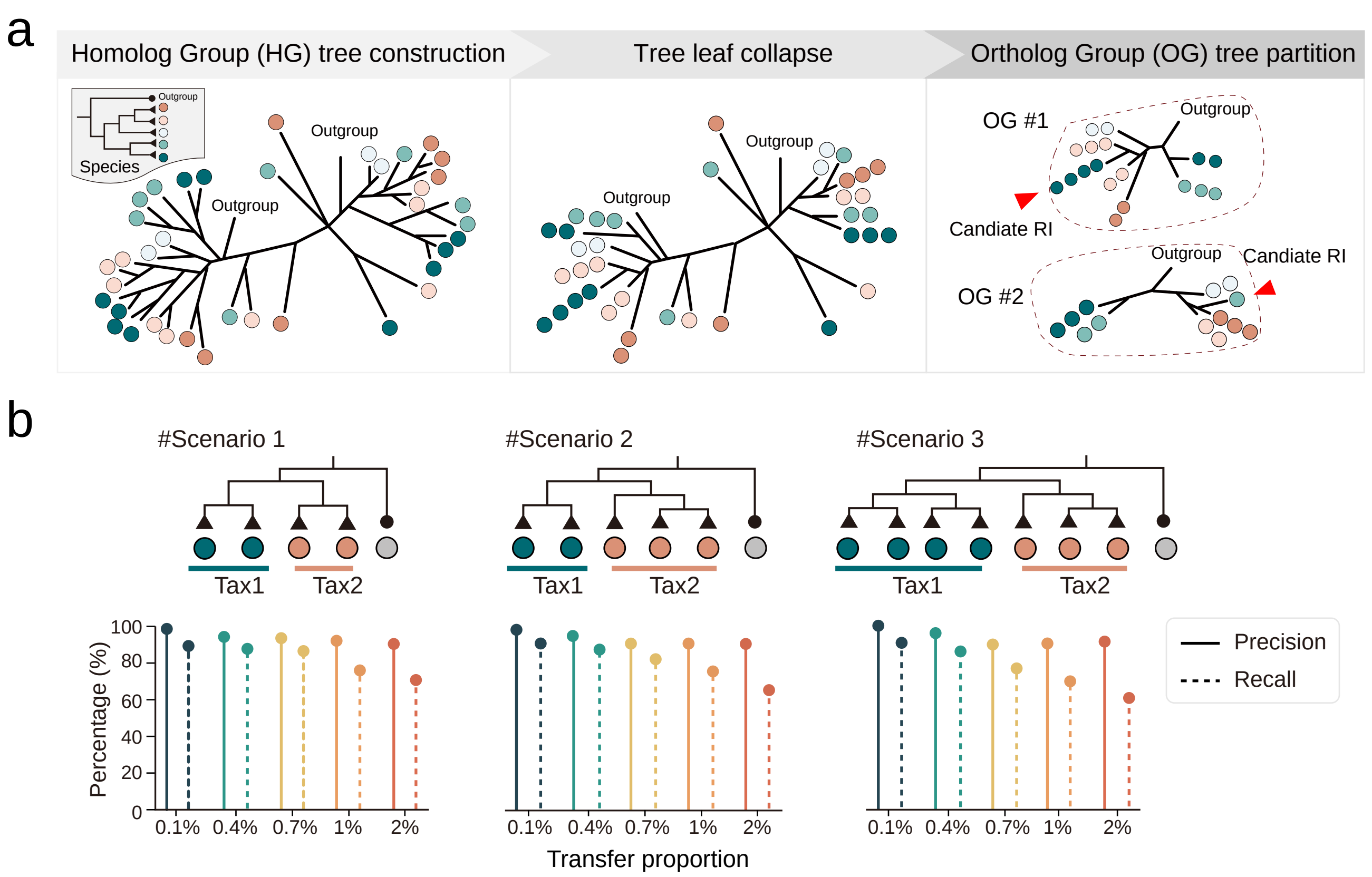

# Figure 2

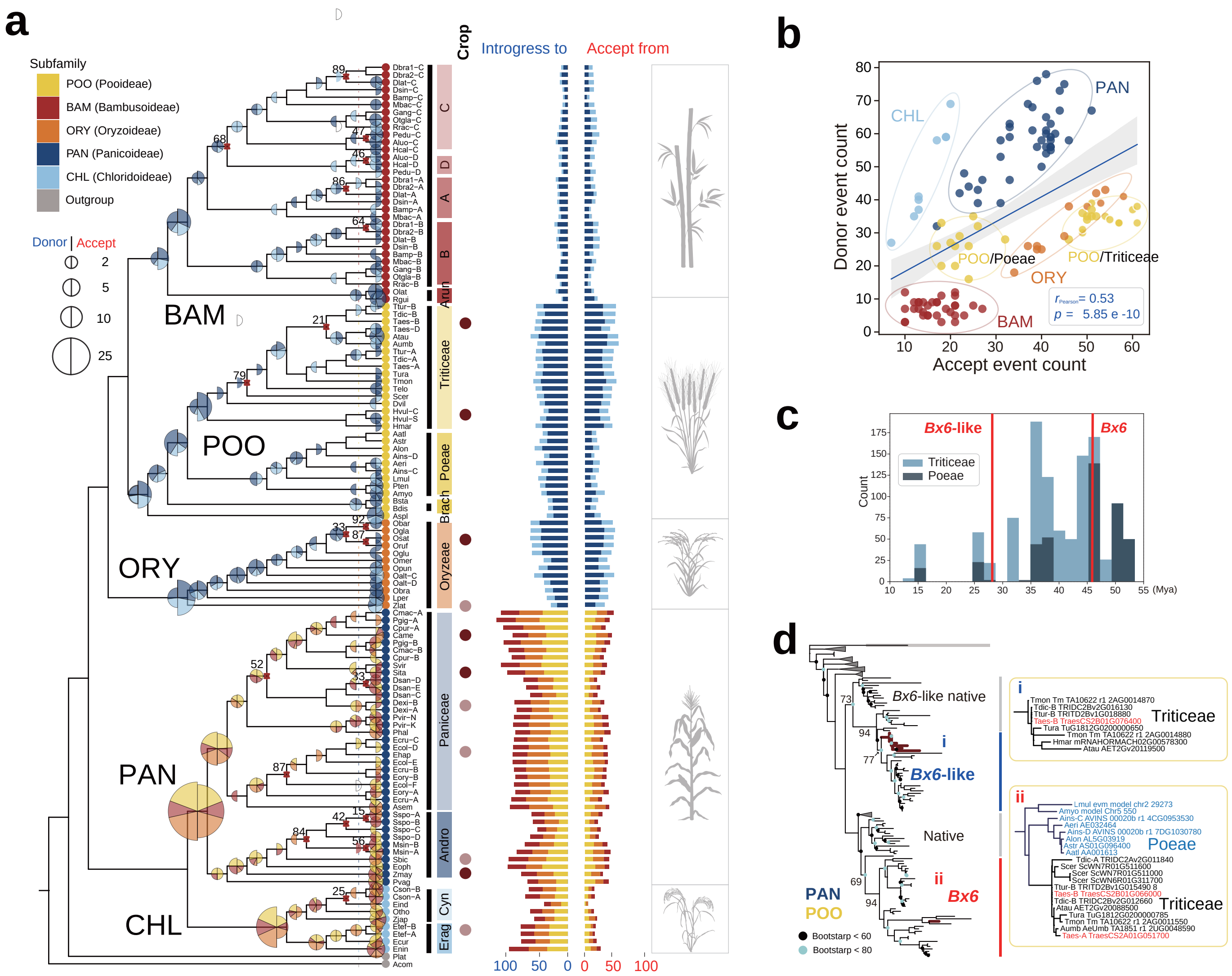

# Figure 3

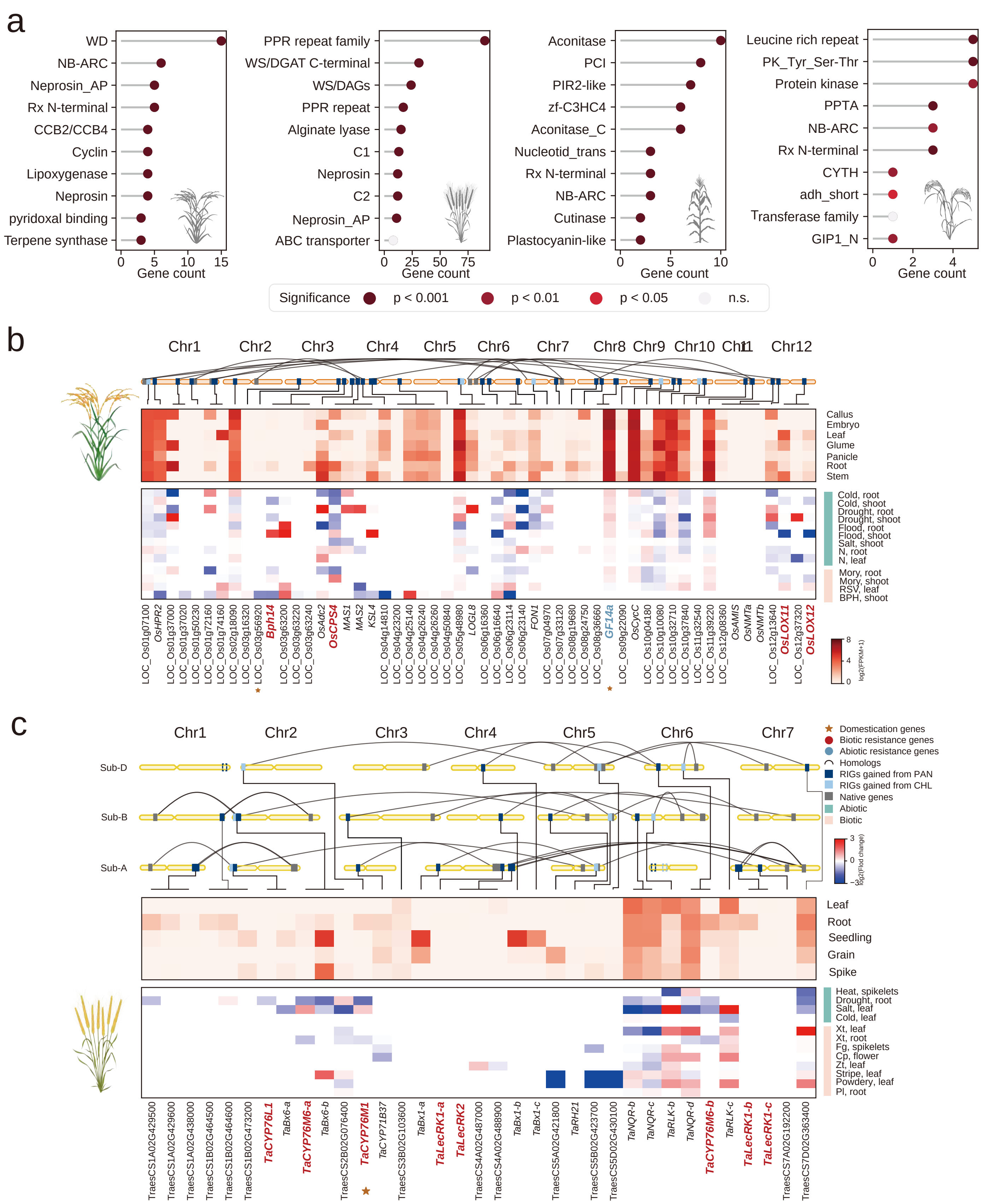

a
WD
NB-ARC
Neprosin_AP
Rx N-terminal
CCB2/CCB4
Cyclin
Lipoxygenase
Neprosin
pyridoxal binding
Terpene synthase
Gene count
PPR repeat family
WS/DGAT C-terminal
WS/DAGs
PPR repeat
Alginate lyase
C1
Neprosin
C2
Neprosin_AP
ABC transporter
Aconitase
PCI
PIR2-like
zf-C3HC4
Aconitase_C
Nucleotid_trans
Rx N-terminal
NB-ARC
Cutinase
Plastocyanin-like
Leucine rich repeat
PK_Tyr_Ser-Thr
Protein kinase
PPTA
NB-ARC
Rx N-terminal
CYTH
adh_short
Transferase family
GIP1_N
Significance p < 0.001 p < 0.01 p < 0.05 n.s.
b
Chr1 Chr2 Chr3 Chr4 Chr5 Chr6 Chr7 Chr8 Chr9 Chr10 Chr11 Chr12
Callus
Embryo
Leaf
Glume
Panicle
Root
Stem
Cold, root
Cold, shoot
Drought, root
Drought, shoot
Flood, root
Flood, shoot
Salt, shoot
N, root
N, leaf
Mory, root
Mory, shoot
RSV, leaf
BPH, shoot
Bph14
OsCPS4
GF14a
OsLOX11
OsLOX12
log2(FPKM+1)
c
Chr1 Chr2 Chr3 Chr4 Chr5 Chr6 Chr7
Sub-D
Sub-B
Sub-A
Domestication genes
Biotic resistance genes
Abiotic resistance genes
Homologs
RIGs gained from PAN
RIGs gained from CHL
Native genes
Abiotic
Biotic
log2(Flod change)
Leaf
Root
Seedling
Grain
Spike
Heat, spikelets
Drought, root
Salt, leaf
Cold, leaf
Xt, leaf
Xt, root
Fg, spikelets
Cp, flower
Zt, leaf
Stripe, leaf
Powdery, leaf
Pl, root
TaCYP76L1
TaCYP76M6-a
TaCYP76M1
TaLecRK1-a
TaLecRK2
TaCYP76M6-b
TaLecRK1-b
TaLecRK1-c

# Figure 4

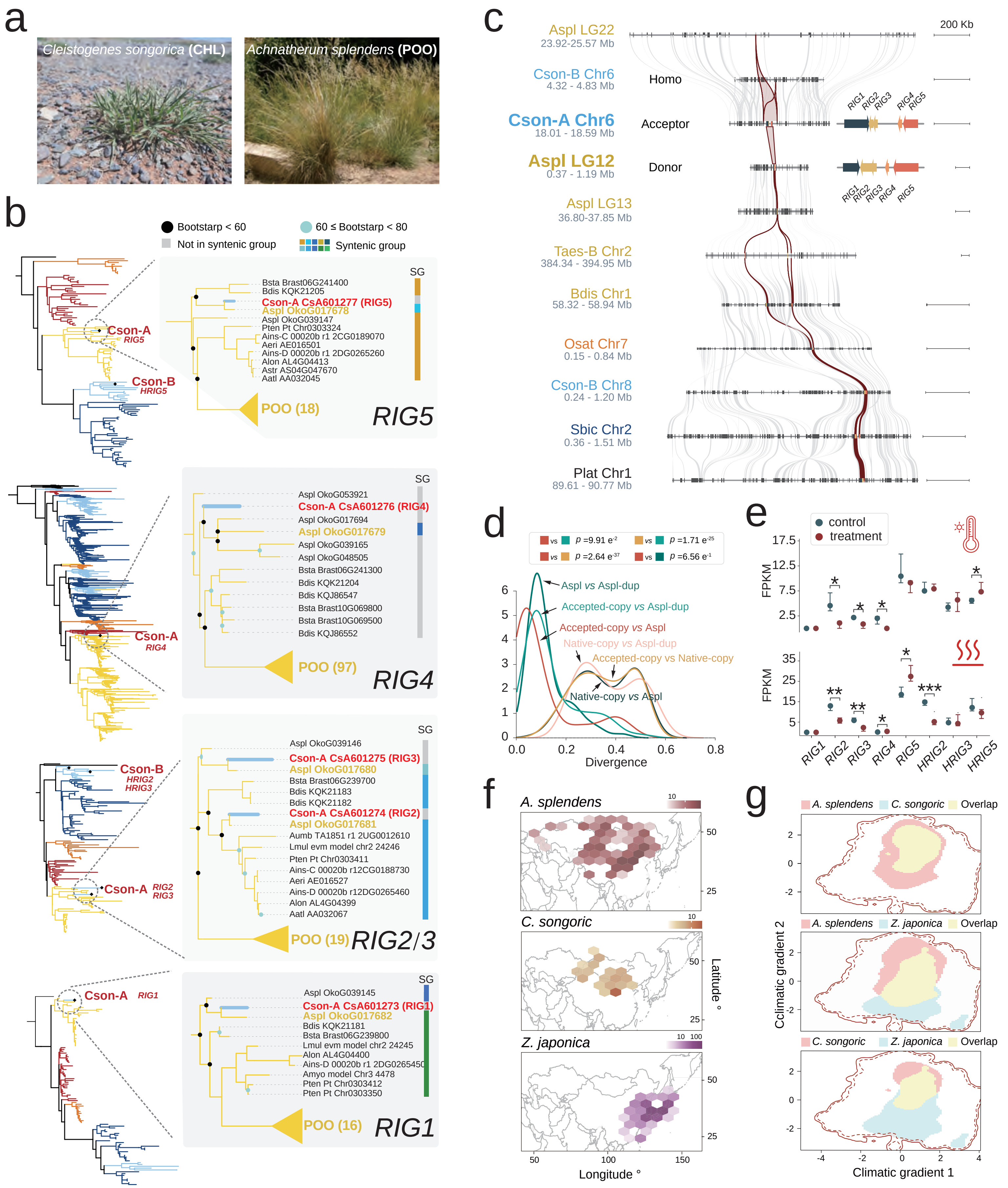

a
Cleistogenes songorica (CHL)
Achnatherum splendens (POO)
b
Bootstarp < 60
60 ≤ Bootstarp < 80
Not in syntenic group
Syntenic group
Cson-A RIG5
Cson-B HRIG5
SG
Bsta Brast06G241400
Bdis KQK21205
Cson-A CsA601277 (RIG5)
Aspl OkoG017678
Aspl OkoG039147
Pten Pt Chr0303324
Ains-C 00020b r1 2CG0189070
Aeri AE016501
Ains-D 00020b r1 2DG0265260
Alon AL4G04413
Astr AS04G047670
Aatl AA032045
POO (18)
RIG5
Cson-A RIG4
Aspl OkoG053921
Cson-A CsA601276 (RIG4)
Aspl OkoG017694
Aspl OkoG017679
Aspl OkoG039165
Aspl OkoG048505
Bsta Brast06G241300
Bdis KQK21204
Bdis KQJ86547
Bsta Brast10G069800
Bsta Brast10G069500
Bdis KQJ86552
POO (97)
RIG4
Cson-B HRIG2 HRIG3
Cson-A RIG2 RIG3
Aspl OkoG039146
Cson-A CsA601275 (RIG3)
Aspl OkoG017680
Bsta Brast06G239700
Bdis KQK21183
Bdis KQK21182
Cson-A CsA601274 (RIG2)
Aspl OkoG017681
Aumb TA1851 r1 2UG0012610
Lmul evm model chr2 24246
Pten Pt Chr0303411
Ains-C 00020b r12CG0188730
Aeri AE016527
Ains-D 00020b r12DG0265460
Alon AL4G04399
Aatl AA032067
POO (19)
RIG2/3
Cson-A RIG1
Aspl OkoG039145
Cson-A CsA601273 (RIG1)
Aspl OkoG017682
Bdis KQK21181
Bsta Brast06G239800
Lmul evm model chr2 24245
Alon AL4G04400
Ains-D 00020b r1 2DG0265450
Amyo model Chr3 4478
Pten Pt Chr0303412
Pten Pt Chr0303350
POO (16)
RIG1
c
200 Kb
Aspl LG22 23.92-25.57 Mb
Cson-B Chr6 4.32 - 4.83 Mb
Homo
Cson-A Chr6 18.01 - 18.59 Mb
Acceptor
Aspl LG12 0.37 - 1.19 Mb
Donor
RIG1 RIG2 RIG3 RIG4 RIG5
Aspl LG13 36.80-37.85 Mb
Taes-B Chr2 384.34 - 394.95 Mb
Bdis Chr1 58.32 - 58.94 Mb
Osat Chr7 0.15 - 0.84 Mb
Cson-B Chr8 0.24 - 1.20 Mb
Sbic Chr2 0.36 - 1.51 Mb
Plat Chr1 89.61 - 90.77 Mb
d
vs p =9.91 e-2
vs p =1.71 e-25
vs p =2.64 e-37
vs p =6.56 e-1
Aspl vs Aspl-dup
Accepted-copy vs Aspl-dup
Accepted-copy vs Aspl
Native-copy vs Aspl-dup
Accepted-copy vs Native-copy
Native-copy vs Aspl
Divergence
e
control
treatment
FPKM
RIG1 RIG2 RIG3 RIG4 RIG5 HRIG2 HRIG3 HRIG5
f
A. splendens
C. songoric
Z. japonica
Longitude °
Latitude °
g
A. splendens
C. songoric
Z. japonica
Overlap
Cclimatic gradient 2
Climatic gradient 1

# Figure 5

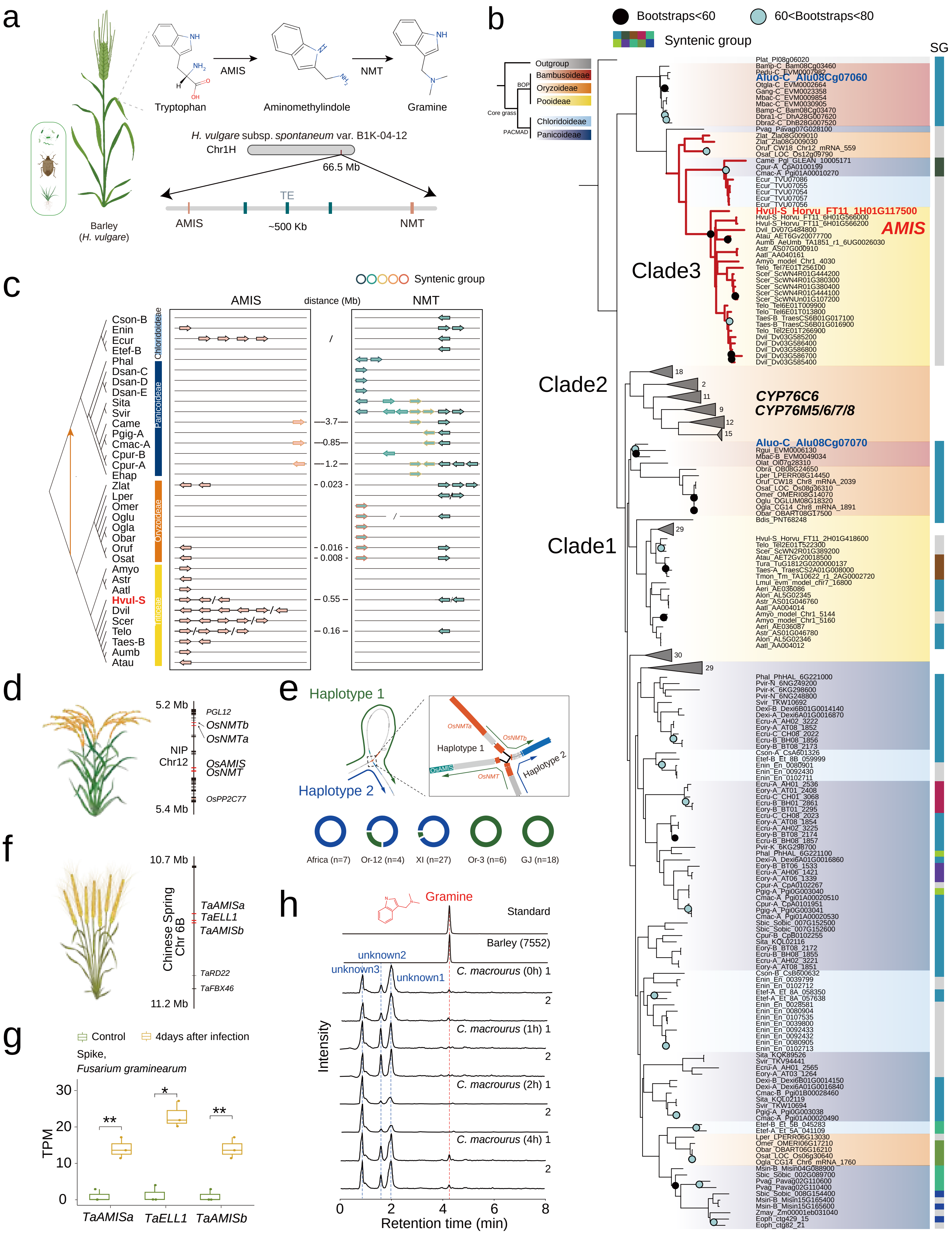